%% file: main.tex
\documentclass[
aps,
prb,
reprint,
superscriptaddress
]
{revtex4-2}

\usepackage[utf8]{inputenc}
\usepackage{siunitx}
\usepackage{graphicx}
\usepackage{amssymb}
\usepackage{booktabs}
\usepackage{color,xcolor}
\usepackage{hyperref}
\usepackage{xspace}
\usepackage{xr-hyper}

\newcommand{\LEmuSR}{LE-$\mu$SR\xspace}
\newcommand{\VSi}{V$\mathrm{_{Si}}$\xspace}

\begin{document}

\title{Device contacts as spin-state selectors for silicon vacancies in 4H‑SiC}




\author{Aurora Teien}
\email{aurora.teien@fys.uio.no}
\affiliation{Department of Physics/ Centre for Materials Science and Nanotechnology, University of Oslo, 0316 Oslo, Norway}

\author{Maria Mendes Martins}
\affiliation{Advanced Power Semiconductor Laboratory, ETH Zürich, Physikstrasse 3, 8092 Zürich, Switzerland}
\affiliation{PSI Center for Neutron and Muon Sciences, 5232 Villigen PSI, Switzerland}

\author{Andreas Gottscholl}
\affiliation{NASA Jet Propulsion Laboratory, California Institute of Technology, Pasadena, CA, USA}

\author{Axel Erlebach}
\affiliation{Advanced Power Semiconductor Laboratory, ETH Zürich, Physikstrasse 3, 8092 Zürich, Switzerland}

\author{Piyush Kumar}
\affiliation{Advanced Power Semiconductor Laboratory, ETH Zürich, Physikstrasse 3, 8092 Zürich, Switzerland}

\author{Erlend Lemva Ousdal}
\affiliation{Department of Physics/ Centre for Materials Science and Nanotechnology, University of Oslo, 0316 Oslo, Norway}

\author{Viktor Bobal}
\affiliation{Department of Physics/ Centre for Materials Science and Nanotechnology, University of Oslo, 0316 Oslo, Norway}

\author{Augustinas Galeckas}
\affiliation{Department of Physics/ Centre for Materials Science and Nanotechnology, University of Oslo, 0316 Oslo, Norway}

\author{Thomas Prokscha}
\affiliation{PSI Center for Neutron and Muon Sciences, 5232 Villigen PSI, Switzerland}

\author{Hannes Kraus}
\affiliation{NASA Jet Propulsion Laboratory, California Institute of Technology, Pasadena, CA, USA}

\author{Corey J. Cochrane}
\affiliation{NASA Jet Propulsion Laboratory, California Institute of Technology, Pasadena, CA, USA}

\author{Ulrike Grossner}
\affiliation{Advanced Power Semiconductor Laboratory, ETH Zürich, Physikstrasse 3, 8092 Zürich, Switzerland}

\author{Lasse Vines}
\affiliation{Department of Physics/ Centre for Materials Science and Nanotechnology, University of Oslo, 0316 Oslo, Norway}

\author{Marianne Etzelm{\"u}ller Bathen}
\email{m.e.bathen@fys.uio.no}
\affiliation{Department of Physics/ Centre for Materials Science and Nanotechnology, University of Oslo, 0316 Oslo, Norway}
\affiliation{Advanced Power Semiconductor Laboratory, ETH Zürich, Physikstrasse 3, 8092 Zürich, Switzerland}

\date{\today}

\begin{abstract}
Optically addressable defect spins in semiconductors are promising building blocks for scalable quantum technologies. Yet, the consequences of conventional contact schemes used in semiconductor device integration for the quantum spin environment remain largely unexplored.
Using the silicon vacancy (V$_\mathrm{Si}$) in silicon carbide (SiC) as a model system, we show that a widely used contact metal, nickel (Ni), intrinsically perturbs the defect spin state and quenches the characteristic emission from the spin-quartet channel of V$_\mathrm{Si}$. Low-energy muon spin rotation further reveals that Ni contacts create a magnetically contaminated region extending at least $\sim$110~nm into the SiC, in stark contrast to non-magnetic contacts such as Ti and Al. 
Moreover, cross-sectional cathodoluminescence measurements on samples with high defect density demonstrate a suppressed quartet-state luminescence from the V$_\mathrm{Si}$ up to 500~nm beneath the Ni contact. This drastic influence on the defect's magnetic environment is accompanied by the appearance of a signature consistent with photoluminescence from the spin-doublet state of V$_\mathrm{Si}$ under the Ni layer, which was absent in other material stacks. These results establish that standard device configurations can drive quantum defects into unwanted charge and spin configurations, underscoring the necessity of precise design to preserve quantum-grade spin environments in semiconductor devices.
\end{abstract}

\maketitle

\section*{Introduction}
Semiconductor defects that combine single-photon and single-spin control, often referred to as color centers or quantum defects, are a critical building block in cutting-edge quantum sensing and communication devices \cite{Weber2010}. 
However, defect opto-spin control is typically reserved for one of the defect's charge states only, highlighting the need for precise charge-state control in quantum devices based on the defect spin-photon interface. Electrical devices such as p-i-n and Schottky diodes offer a precise means towards volume-dependent preferential charge-state occupation \cite{bathen_manipulating_2021}, but require mature and reliable semiconductor processing \cite{streetman2000solid}. 
Silicon carbide (SiC) is emerging as an attractive platform for future quantum devices, given its industry-grade material and processing \cite{Kimoto_2014}, unparalleled by other quantum defect hosts besides silicon, paired with varied color center emission spectrum ranging from the visible to telecom wavelengths \cite{castelletto_silicon_2020, zhang_material_2020}. 

Among the  prospecting color centers in SiC we find the silicon vacancy (V$_\mathrm{Si}$) with indistinguishable single-photon emission \cite{Morioka_2020} in the near infrared, paired with spin control \cite{kraus_roomtemp_2014,widmann_coherent_2015},  in the negative charge state V$_\mathrm{Si}^-$ \cite{sorman_silicon_2000, janzen_silicon_2009}. 
The V$_\mathrm{Si}^-$ also shows a high sensitivity to local changes in the magnetic environment \cite{kraus_magnetic_2014}, and is attractive for, e.g., quantum sensing of various physical properties \cite{cochrane_vectorized_2016, ohshima_creation_2018, soykal_quantum_2017, anisimov_optical_2016}. 
In the ground state, V$_\mathrm{Si}^-$ exhibits a spin-quartet $S=\frac{3}{2}$ spin manifold with a zero-field splitting of $\sim$70~MHz between $m_s=\pm\frac{1}{2}$ and $m_s=\pm\frac{3}{2}$ \cite{ohshima_creation_2018, soykal_silicon_2016}, where the opto-spin evolution of the color center is typically read out from the quartet spin channel.

Although the quantum-compatible properties of the V$_\mathrm{Si}^-$ are read-out from the singly negative charge state, this configuration is not preferred in typical n- or p-doped samples \cite{Hornos2011}, instead necessitating semi-insulating material for optimal charge-state occupation. Viable charge-state control strategies include Fermi-level design at the material level \cite{son_charge_2021}, optical control via dual-laser excitation \cite{Wolfowicz2017}, and electrical device driving via Schottky barrier diodes (SBDs) \cite{bathen_electrical_2019} or p-i-n diodes \cite{widmann_2019}. 
However, the precise band-bending imposed by different contact metals and its impact on the defect charge state, combined with possible adverse effects of the metal-semiconductor interface on the color center's charge, spin and optical stability, remain an open question.

Here, we compare the influence of commonly used Schottky contact materials on 4H-SiC, 
namely Al, Ti, and Ni, on the spin and optical state stability of quantum defects, using the V$_\mathrm{Si}$ as a test case. 
An additional contact material, Pd, is also considered herein. 
The V$_\mathrm{Si}$ emission stability is evaluated using optically detected magnetic resonance (ODMR) and photoluminescence (PL) spectroscopy, while low-energy muon spin rotation (\LEmuSR) and cross-sectional cathodoluminescence (CL) spectroscopy offer valuable insights into the evolution of the defect signal at the contact-semiconductor interface and further into the material. The experimental techniques are complemented by device simulations to provide the band bending profile of each Schottky diode type and predicted charge-state occupation of the Si vacancy. 
We find that Ni compromises the magnetic environment of quantum defects, fully quenching both the optical and spin signature of the V$_\mathrm{Si}^-$ up to $\sim500$~nm into the SiC. 
A photoluminescence signature appearing where the doublet state of V$_\mathrm{Si}^-$ is expected to emit evidences the involvement of spin-state mixing between the doublet and quartet states. 
Our findings emphasize the importance of controlling each fabrication step when designing quantum devices, and understanding the possible causes of spin-state mixing in solid-state color centers.

 \begin{figure*}[t]
    \centering
    \includegraphics[width=0.9\linewidth]{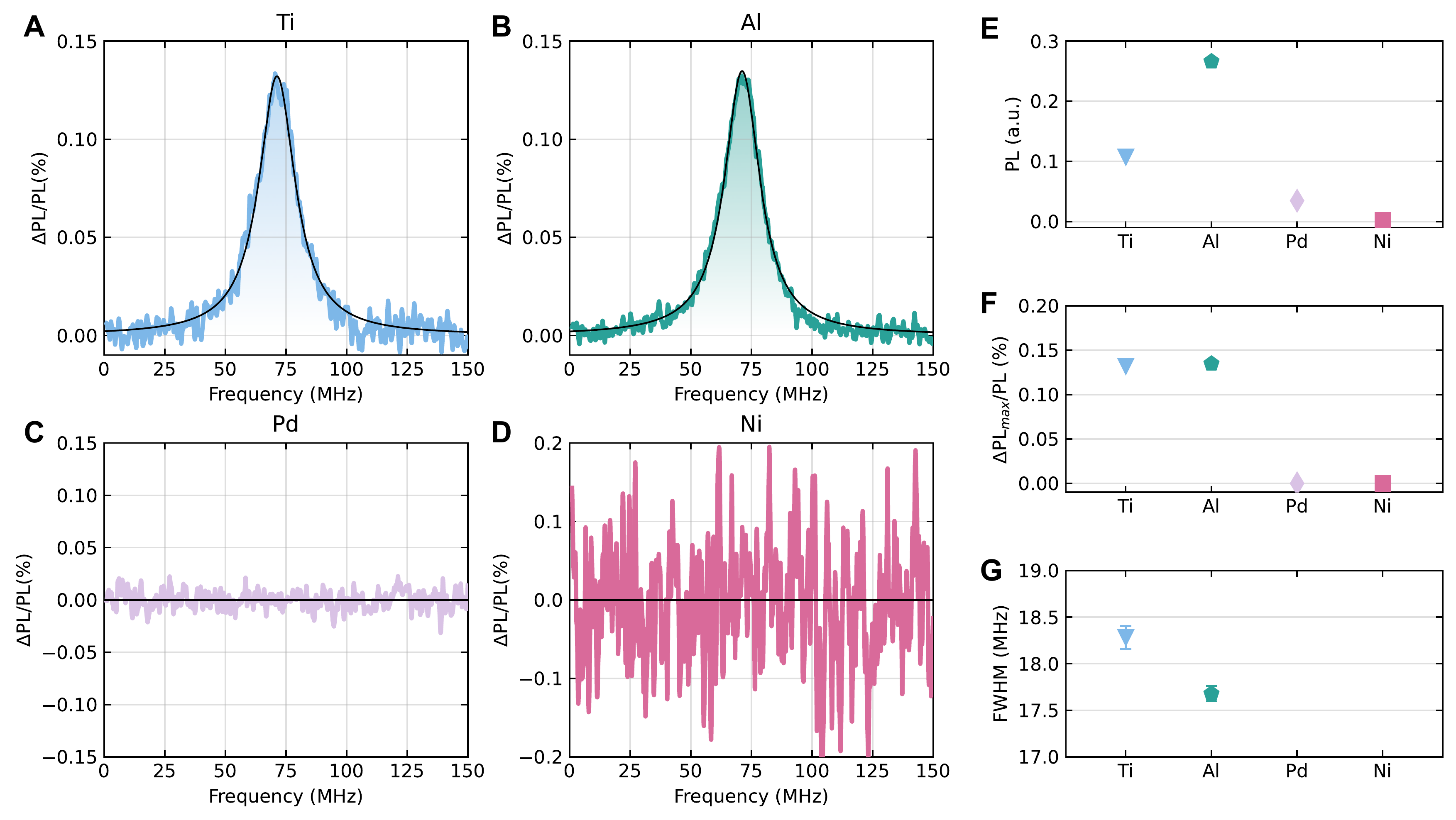}
    \caption{\textbf{Defect magneto-optical response to metallization.} Room-temperature ODMR ($\Delta$PL/PL) signal from 4H-SiC  epi-layers having a  V$_\mathrm{Si}$-concentration of $\sim$\SI{1e16}{\per\centi\meter\cubed} within a $\sim$\SI{0.6}{\micro\meter} wide defective box profile, and 100~nm thick contacts of (A) Ti, (B) Al, (C) Pd and (D) Ni, collected in a back-side geometry. Only the Ti and Al samples exhibit a pronounced ODMR signal, with $\Delta$PL/PL$> 0.01\%$. The remaining panels show (E) photoluminescence, (F) maximum ODMR contrast, and (G) ODMR linewidth (full width at half maximum, FWHM) collected for the four samples.
}
    \label{fig:PL_ODMR}
\end{figure*}

\section*{Results} 

Color-center-based quantum technologies commonly rely on optically detected magnetic resonance (ODMR), for which the contrast, i.e., the relative change in photoluminescence (PL) intensity, depends critically on the color center occupying the appropriate optically active charge state. For silicon vacancies V$_\mathrm{Si}$ in 4H-SiC, the negatively charged V$_\mathrm{Si}^-$ state is optically bright and can be probed by monitoring its PL emission (here: 850–1000~nm, see Methods) while sweeping the radio frequency across the spin resonances. When the oscillating field is resonant with a ground-state spin transition ($\sim$70 MHz for V$_\mathrm{Si}^-$ V2), it drives spin flips that redistribute the ground-state populations. Spin-dependent relaxation and optical cycling then convert this redistribution into a change in the detected PL intensity~\cite{liu2024silicon}. Observation of an ODMR resonance therefore provides a direct spectroscopic signature of the optically active (V$_\mathrm{Si}^{-}$) charge state. The ODMR linewidth further probes the local spin environment: broadening beyond the intrinsic linewidth can arise from spatial variations in the local magnetic field and other sources of inhomogeneous broadening that distribute the spin-transition frequencies.

Fig.~\ref{fig:PL_ODMR} shows the room-temperature ODMR response from 4H-SiC epi-layers having a V$_\mathrm{Si}$-concentration of $\sim$\SI{1e16}{\per\centi\meter\cubed} within a $\sim$\SI{0.6}{\micro\meter} wide defective box profile, and 100~nm thick contacts of (A) Ti, (B) Al, (C) Pd and (D) Ni, collected in a back-side geometry. Pronounced ODMR resonances are observed only for the Ti- and Al- contacted samples, both centered around the expected zero-field splitting (ZFS) of $\sim$70~MHz for \VSi. The ODMR linewidths are comparable for the two samples, with only a marginal broadening for Ti relative to Al (see also Table~\ref{tab:contactStats}). In contrast, the Ni contact sample shows neither a measurable PL nor an ODMR signal, resulting only in a noise-limited spectrum, indicating that the spin-dependent optical response is fully suppressed. Finally, the Pd contact sample exhibits finite (albeit weak) PL, but no detectable ODMR contrast above the measurement noise floor

The integrated PL recorded during the ODMR measurements is summarized in Fig.~\ref{fig:PL_ODMR}(E). 
The Al contact exhibits the highest integrated PL intensity, followed by Ti, whereas the PL is strongly reduced for Pd and completely quenched for Ni. Figs.~\ref{fig:PL_ODMR}(F)-(G) summarize the ODMR contrast and linewidth, respectively, exhibiting a similar trend, but with closer agreement between Ti and Al for the ODMR contrast. Table~\ref{tab:overview_contacts} provides an overview of the contact-dependent results obtained from ensemble PL and ODMR measurements as well as capacitance-voltage measurements to determine the Schottky barrier height obtained for each contact metal.

\begin{table*}[t]
    \centering
    \caption{\textbf{Schottky diode details.} Summary of Schottky contact details including results from PL and ODMR measurements on the samples shown in  Fig.~\ref{fig:PL_ODMR} (i.e., summarizing in which samples PL and ODMR signal was collected), the measured ODMR linewidths $\Delta$f, expected magnetization of each contact metal, their molar magnetic susceptibility ($\chi_M$), and the Schottky barrier height $\Phi_B$ as extracted from room-temperature capacitance-voltage measurements. } \label{tab:overview_contacts}
    \begin{tabular*}{\textwidth}{@{\extracolsep\fill}lcccccc }
    \toprule%
        Metal & PL   & ODMR  & Linewidth $\Delta$f (MHz) & Magnetization & $\chi_M$ (cm$^3$mol$^{-1}$ )& $\Phi_B$ (eV)  \\
        \midrule 
        Ti  & x & x & 18.3 & Low  & $1.51\times10^{-4}$ \cite{collings_magnetic_1970, haynes_crc_2015}& 1.45  \\
        Al  & x & x & 17.7 & Low  & $1.65\times10^{-5}$ \cite{haynes_crc_2015}& 0.72  \\
        Pd  & x & - & - & Low  &$5.40\times10^{-4}$ \cite{jamieson_magnetic_1972, haynes_crc_2015}& 2.20 \\
        Ni  & - & - & - & High  & Ferromagnetic  \cite{danan_new_1968} & 1.88  \\
    \botrule
    \label{tab:contactStats}
    \end{tabular*}
\end{table*}

\begin{figure*}[t]
    \centering
    \includegraphics[width=\linewidth]{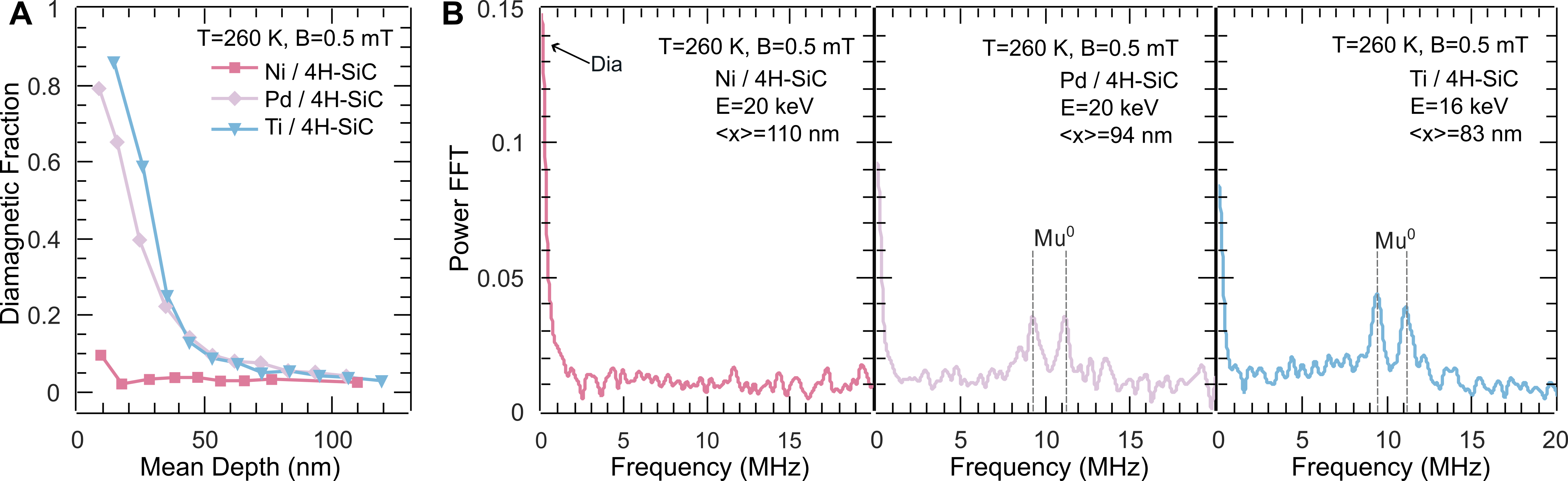}  
        \caption{\textbf{Magnetic environment imposed by contact metal.} (A) Diamagnetic fraction as a function of mean depth measured at \SI{260}{\kelvin} and \SI{10}{\milli\tesla}, measured for 4H-SiC epi-layers exposed to 1.8~MeV protons to a fluence of \SI{1e14}{\per\centi\meter\cubed} and having a \SI{20}{\nano\meter} top layer of Ni, Pd or Ti. (B) Fourier transform of the measured time spectra under \SI{0.5}{\milli\tesla} and \SI{260}{\kelvin}, revealing that only the diamagnetic component is observable in the Ni/SiC stack. While in the samples with Pd and Ti contacts the paramagnetic state Mu$^0$ can be seen.}

    \label{fig:uSR_results}
\end{figure*}

\subsection*{Magnetic environment }
To investigate the mechanism for strong emission quenching observed for the \VSi under Ni- and Pd-contacts in Fig.~\ref{fig:PL_ODMR}, the influence of metallization on the magnetic environment near the surface of 4H-SiC was determined using LE-$\mu$SR measurements. 
In LE-$\mu$SR, the muon ($\mu^+$) serves as a local sensor to the magnetic and electronic environment experienced by a uniform ensemble of Si vacancies in the top $\sim150$~nm of the SiC. The samples were prepared on n-type 4H-SiC epitaxial layers with doping concentration of \SI{2.8e15}{\per\cubic\centi\meter}, and Si vacancies were created by 1.8~MeV proton irradiation to a fluence of \SI{1e14}{\per\square\centi\meter} to ensure detection of \VSi with LE-$\mu$SR \cite{Woerle_2020}. Note that the entire region probed by \LEmuSR, extending from the near surface to a depth of $\sim110$~nm, contains a homogeneous \VSi density.

In the LE-$\mu$SR experiment, a fraction of the implanted muons remain as isolated positively charged muons ($\mu^+$), giving rise to the diamagnetic fraction. Muons may also capture an electron and form muonium (Mu$^0$). 
Since defects can act as trapping centers for electrons and stabilize defect-bound muonium states, the measured diamagnetic fraction provides a sensitive probe of defect populations in semiconductors.
Particularly in SiC, \VSi leads to a reduction of the diamagnetic fraction, as the muons form a neutral state in a Mu$^0$-\VSi complex \cite{Woerle_2020}.

The diamagnetic fractions measured for samples with Ni, Pd, and Ti contacts are shown as a function of mean depth from the surface in Fig.~\ref{fig:uSR_results}(A). In the Ni sample, which is ferromagnetic at temperatures below 627~K, strong internal magnetic fields lead to rapid depolarization of the muon spin, effectively suppressing the observable fraction within the experimental time window. As a result, the measured diamagnetic fraction is significantly reduced. 
In contrast, Pd and Ti are non-magnetic under these conditions, so the muon spin polarization relaxes substantially slower, yielding a larger observable fraction, albeit slightly higher in Ti than Pd. Consequently, the diamagnetic fraction can be interpreted as a measure of the non-magnetic fraction, which is substantially higher in Pd and Ti than in Ni. For comparison, previous works have included material stacks of Al and SiC \cite{Kumar_2023}, yielding no measurable impact on the muon-spin depolarization in SiC, similar to that observed for Ti and Pd. 

In the SiC layer, the low diamagnetic fraction observed in Fig.~\ref{fig:uSR_results}(A) is consistent with previous observations in samples containing \VSi defects \cite{Woerle_2020}.
However, for the Pd and Ti samples,  the diamagnetic fraction does not drop immediately below the metal–SiC interface, which can be attributed to the signal containing contributions from both the metal and the SiC layer (illustrated in Supplementary Methods Fig.~S3).
Notably, in the sample with Ni, the diamagnetic fraction is even lower, approaching the lower detection limit of \LEmuSR. This evidences that stray or inhomogeneous magnetic fields originating from the Ni layer can penetrate into the adjacent SiC region, causing additional depolarization of the muon-spin within the SiC. As a result, even muons that stop at $\sim$\SI{110}{\nano\meter} from the interface may experience sufficiently strong or fluctuating local fields to reduce the observable diamagnetic fraction.

Fig.~\ref{fig:uSR_results}(B) shows the Fourier transform of the time spectra measured at \SI{0.5}{\milli\tesla} and \SI{260}{\kelvin} for implantation depths between 80 and 110~nm. A low-frequency feature corresponding to the diamagnetic component is observed in all samples. However, the Mu$^0$ signal is absent when the Ni layer is present. This finding is consistent with stray magnetic fields from the Ni layer penetrating into the SiC and substantially increasing the depolarization rate of the paramagnetic Mu$^0$ state. Consequently, the Mu$^0$ signal becomes unresolvable within the present experiment.

The complete suppression of the ODMR response in the Ni sample and the absence of spin contrast in the Pd sample demonstrate that the spin properties of the \VSi\ are strongly affected by the contact material. However, these ODMR measurements provide only an averaged response over the entire implanted defect layer, while \LEmuSR probes the near-surface region. Thus, the results do not reveal how far the contact-induced perturbation extends into the SiC, nor possible variations throughout the SiC. To address this question, cross-sectional CL measurements with a spatial resolution of $\sim100$~nm were performed.

\subsection*{Mapping the near-surface defect response }

\begin{figure*}[t]
    \centering
    \includegraphics[width=0.9\linewidth]{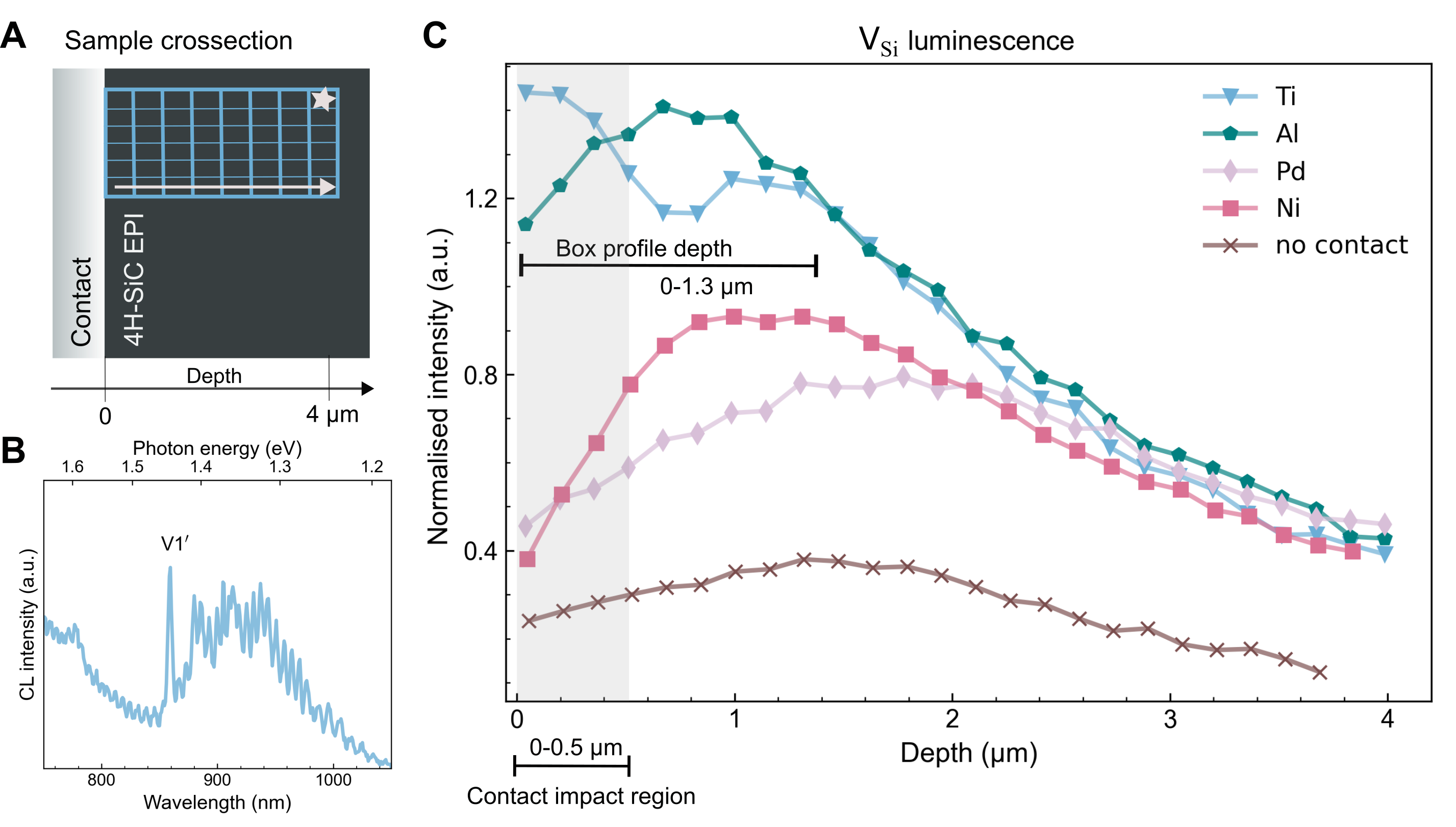}
    \caption{\textbf{Mapping the near-surface defect response to electro-magnetic environment.} (A) Schematic drawing of the sample cross-section with an arbitrary 2D measurement grid. The thickness of the contact and the pixel sizes are exaggerated for illustrative purposes. Each pixel (grid point marked with a star) contains a single luminescence spectrum, as seen in (B). Mapping the height of the V1$^{\prime}$ peak from the V$_\mathrm{Si}$ defect for each pixel in a row (visualised as an arrow in (A)) yields the depth profile shown in (C). This depth profile shows the evolution of the V$_\mathrm{Si}$-luminescence under Ti, Al, Pd, and Ni contacts. The luminescence signal is normalised to the background signal. 
    The measurement is performed at 80~K, with probe current 1~nA, acceleration voltage 10~kV and pixel size 150~nm.}
    \label{fig:CLS0}
\end{figure*}

For this experiment, a new type of n-type 4H-SiC sample with a $\sim$\SI{1.3}{\micro\meter} box profile of silicon vacancies (\VSi) to a density of  $5\times10^{15}$~\SI{}{\per\centi\meter\cubed} was fabricated (for further details, see Supplementary Methods). Each sample is equipped with spatially separated contacts of all of the previously studied materials Ti, Al, Pd, and Ni, deposited sequentially using electron-beam evaporation through a shadow mask. Since the area of interest lies underneath the contacts, cross-sections of the samples were fabricated via laser cutting and cleaving, and used for CL-inspection as shown schematically in Fig.~\ref{fig:CLS0}(A). An example of a resulting luminescence spectrum in a random electron beam spot is shown in Fig.~\ref{fig:CLS0}(B), where the \VSi ZPL is marked V1$^{\prime}$. 

The defect luminescence signal observed in CL is expected to be proportional to the defect concentration, provided the defect density remains below saturation, and thus a higher emission intensity is expected in the box profile region than in the tail. Fig.~\ref{fig:CLS0}(C) shows the normalised CL intensity from the V$_\mathrm{Si}$ as a function of depth from the surface for sample regions situated beneath the four different contact metals: Ti (blue curve), Al (green curve), Pd (light purple curve) and Ni (pink curve). The CL signal from a region without any contact is included for reference (brown curve), with a substantially lower integrated CL intensity than from beneath any of the contact metals.  
As expected, the \VSi emission intensity is found to decrease 
deep into the SiC substrate while being stronger closer to the metal-semiconductor interface. 
The box-profile is simulated to extend $\sim$\SI{1.3}{\micro\meter} into the sample from the surface (see Supplementary Methods), which is in good agreement with the depth where the CL-intensity starts decreasing in Fig.~\ref{fig:CLS0}(C) for all contacts. However, \VSi-related signatures are still clearly visible several micrometers into the sample, evidencing that damage is generated even beyond the projected range of the box-profile region. 
A common feature of all CL depth profiles collected from contacted samples in Fig.~\ref{fig:CLS0}(C) is that they converge and follow the same slope at depths $\sim$2.5-\SI{4}{\micro\meter}. In this region, the high \VSi-concentration from the box profile drops off and the effects of the Schottky contact no longer impact the defect luminescence. The slope originates from the Gaussian tails of the \VSi-box-profile, combined with carrier diffusion from the excitation region, since the CL signal at a given observation point is proportional to the local radiative recombination rate. 

Fig.~\ref{fig:CLS0}(C) reveals large differences in emission from the $\mathrm{V_{Si}}^-$ at shallow depths when comparing the different metals. In the top $\sim$\SI{0.6}{\micro\meter} from the surface, marked by the gray box in Fig.~\ref{fig:CLS0}(C), the luminescence under the Ni and Pd contacts is almost completely quenched. This is in stark contrast to the case for Al and Ti, as both exhibit maximal (Ti) or close to maximal (Al) V1$^{\prime}$ emission relative to the background CL response near the metal-SiC interface. 
The quenching effect is especially pronounced for the Ni contact, as the V1$^{\prime}$ emission increases by a factor of $\sim2.5\times$ from the contact-SiC interface to a maximal intensity after $\sim$\SI{0.8}{\micro\meter}. In comparison, the normalised luminescence under the Ti contact is $\sim4\times$ higher than beneath the Ni contact at a depth of 150~nm from the surface, despite being collected from different locations on the same sample. These results evidence that the choice of metal-semiconductor interface drastically impacts color center emission, with the quenching effect imposed by Ni extending several hundred nanometer into the active device region for the test case of the V$_\mathrm{Si}$. In comparison, Pd inflicts a weaker absolute reduction of the \VSi emission intensity, but extending even further into the SiC.

\begin{figure*}[t]
    \centering
    \includegraphics[width=0.9\linewidth]{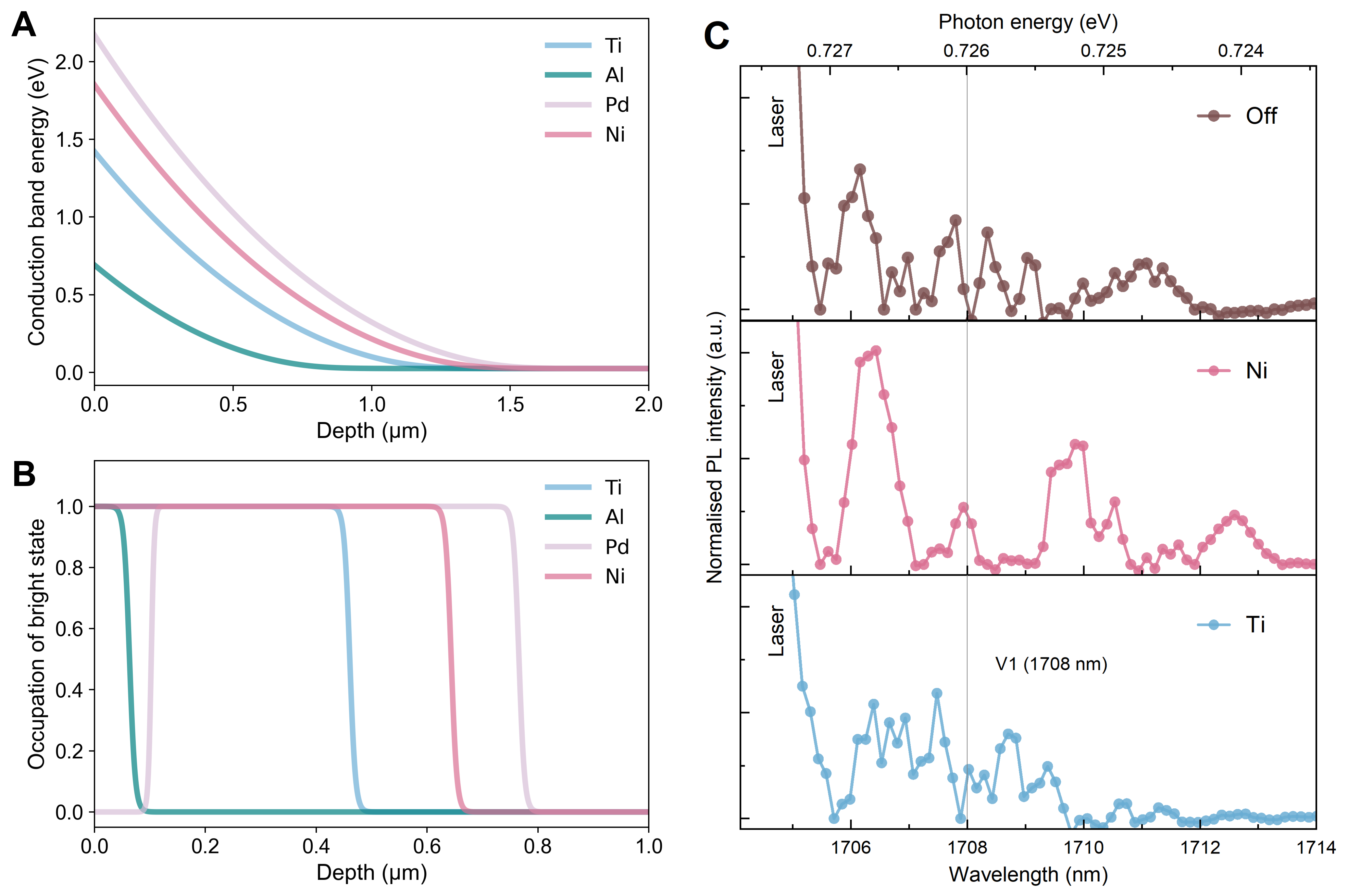}
        \caption{\textbf{Impact of metallization on band bending, charge-state occupation and spin-state mixing.} (A) Simulated conduction band energy as a function of depth into the material for the different contact metals.
        (B) Device simulation of the singly negative charge state occupation of the \VSi for the contact materials Ti, Al, Pd, and Ni. The simulation was performed using a multi-state acceptor trap with energies of 1.9~eV (-) and 0.6~eV (2-) from the conduction band edge, a defect concentration of \SI{5e13}{\per\centi\meter\cubed}, at a temperature of 80~K. 
        (C) Normalised PL intensity under different contacts (Ni and Ti) compared to areas without contacts (brown) at \SIrange{1704}{1714}{\nano\meter}. PL measurements are performed with a 852~nm excitation source at temperature 3.6~K.}
    \label{fig:TCAD_PL_fig}
\end{figure*}

The ODMR, PL, \LEmuSR and CL measurements all reveal a similar effect; a substantial variation in \VSi opto-spin properties depending on the device design. Indeed, where Ti/SiC and Al/SiC structures substantially enhance \VSi emission intensities as compared to the contact-free case, and maintain strong ODMR contrast with a narrow linewidth, Pd and Ni cause a dramatic emission intensity reduction in the top $\sim500$~nm of the active region of the simple quantum device studied herein. 
Two alternative hypotheses are proposed to explain the differences in depth-dependent \VSi\ emission: (i) that the band bending imposed by the Schottky diode nature of the contact promotes selective occupation of dark charge states, or (ii) that an environmental contamination from the deposited metal, likely a magnetic field according to the LE-$\mu$SR measurements, penetrates into the SiC, akin to a proximity effect in superconducting hybrid structures, and drives the \VSi out of an emitting state.

\subsection*{Simulated band bending for charge-state control}
We start by evaluating the first of these hypotheses by simulating each Schottky diode device to determine the band bending imposed on the SiC from different contact metals and its effect on the V$_\mathrm{Si}$ charge state occupation.

Electrical devices can be used to enhance emission output and selectively drive color centers between bright and dark emission regimes, as seen previously for, e.g., the \VSi\ \cite{bathen_electrical_2019,widmann_2019,Bathen_2023} and divacancy \cite{anderson_2019} in 4H-SiC. This is interpreted as a preferential occupation of the bright charge state over dark ones due to band bending, where the dark states are either neutral or doubly/triply negative for the case of the \VSi\ \cite{Hornos2011,bathen_electrical_2019}. However, the slope of the band bending, and thus the depth distribution of defect charge state occupation, depends on the work function and thus the Schottky barrier of the contact material. This can impact the maximum normalised CL intensities for the different contacts shown in Fig.~\ref{fig:CLS0}(C). 

The simulated conduction band bending of the different Schottky metal contacts is shown in Fig.~\ref{fig:TCAD_PL_fig}(A). 
The simulations were performed in equilibrium (no applied bias voltage) and at a temperature of 80~K for the four different contact materials Al, Ti, Ni, and Pd.
Acceptor traps with two charge states at energies of 0.6~eV and 1.9~eV below the conduction band edge are included in the simulation to emulate the $(-/2-)$ and $(0/-)$ bright--to--dark charge-state transitions of the \VSi\ \cite{Hornos2011,bathen_electrical_2019}, respectively. We employ a \VSi defect concentration of $5\times10^{13}$~\SI{}{\per\centi\meter\cubed} to ensure that the epi-layer doping ($N_D=$\SI{1e15}{\per\centi\meter\cubed} in the model) is not compensated, thus preserving the full band bending properties of the diode. 
The predicted defect concentration in the samples used for the CL measurements, $\sim 5\times10^{15}$~\SI{}{\per\centi\meter\cubed}, is found to cause full compensation of charge carriers during simulations. 
The lower defect concentration simulations employed here are expected to provide a closer depiction of the experimental conditions than the fully compensated case, as charge carriers are generated by the excitation in both PL, CL and ODMR measurements, effectively re-establishing a conductive region close to the contact. This is particularly true for the CL measurements, where a total acquisition time of $\sim35$~min for each depth profile in Fig.~\ref{fig:CLS0}(C) yields substantial electron-hole pair generation within the probed region. Simulations employing the higher defect concentrations are included in Supplementary Results for comparison. 

Fig.~\ref{fig:TCAD_PL_fig}(A)-(B) shows the simulated conduction band energy and bright charge state occupation, respectively, as a function of depth from the metal-semiconductor interface for each contact metal. 
For the low \VSi density of $5\times10^{13}$~\SI{}{\per\centi\meter\cubed}, the occupation of the defect energy states does not influence the electrostatics. The band bending is determined by the metal contact and the doping concentration only. 
In the simulation results shown in Fig.~\ref{fig:TCAD_PL_fig}(B), the highest occupation of the singly negative charge state of the \VSi in the surface region occurs under the metal contacts Al, Ti, and Ni. However, Al drops off quickly, and after some tens of nm from the metal-semiconductor interface, the \VSi transitions from $\mathrm{V_{Si}}^-$ to the doubly negative charge state.
In contrast, the Ti and Ni contacts promote selective occupation of $\mathrm{V_{Si}}^-$ for the top $\sim400$~nm and $750$~nm of the SiC region, respectively. 

Interestingly, Pd exhibits low initial occupation close to the metal-semiconductor interface, caused by the occupation of the dark neutral charge state due to the large Schottky barrier, followed by a large region of $\sim700$~nm where the $\mathrm{V_{Si}}^-$ is preferred. Indeed, this evidences that the initial dip in CL intensity under the Pd contact in Fig.~\ref{fig:CLS0}(C) can be explained by band bending effects. For the high defect concentration simulations (see Supplementary Figure S10), the bright state occupation is low throughout the entire region, and increasing from zero only after $\sim$500~nm. Thus, both simulated pictures are in keeping with the experimental observations, with an overall lower emission intensity found by both PL, CL and ODMR under Pd contacts as compared to Ti and Al. 

In contrast, under the Ni contact, the \VSi should according to the simulations in Fig.~\ref{fig:TCAD_PL_fig}(B) be occupying the singly negative charge state throughout the entire surface region and extending $\sim700$~nm into the material. Even for the high defect density case with full carrier compensation (see Supplementary Figure S10), the \VSi emission should peak close to the surface and then drop off. This is in stark contrast to the observed behavior, where the \VSi emission experiences substantial quenching in the near-surface region under the Ni layer.

\subsection*{ZPL dependence on spin channel} 
As evidenced above, the band bending effects alone are not sufficient to explain the full quenching effect of the \VSi\ under Ni contact metal. Our second hypothesis takes the spontaneous magnetization of Ni into account. Recent work on the electronic structure of the $\mathrm{V_{Si}}^-$ \cite{younesi_unraveling_2026} reveals that altering the $\mathrm{V_{Si}}^-$'s spin channel from the quartet state, $S=\frac{3}{2}$, to the doublet state, $S=\frac{1}{2}$, yields a new set of transition energy levels. Thus, the well-known quartet ZPLs of the $\mathrm{V_{Si}}^-$, i.e., V1, V1' and V2, will no longer be visible. Instead, the doublet channel of the $\mathrm{V_{Si}}^-$ contains multiple possible transition pathways, where ZPLs were measured at 1132~nm, 1137~nm, and 1708~nm for $\mathrm{V_{Si}}^-$(\textit{h}), and at 1158~nm, 1168~nm, and 1787~nm for $\mathrm{V_{Si}}^-$(\textit{k}) \cite{younesi_unraveling_2026}.

PL measurements of the same samples that were used for the CL measurements, but now from a backside geometry to isolate emission from beneath the different contact metals, reveal that a new emission line appears at $1708$~nm that is not present under the Ti-contact nor in the reference sample (see Fig.~\ref{fig:TCAD_PL_fig}(C)). 
This is precisely the spectral location of the lowest energy doublet transition of $\mathrm{V_{Si}}^-(h)$ found in Ref.~\cite{younesi_unraveling_2026}. Note that of the listed possible doublet transitions, 1787~nm is outside the detection range of our detector, while the 1132~nm luminescence is overlapping with the PL1 emission from the $hh$-configuration of the divacancy (see Supplementary Figure S7). 
It should be noted that the detector efficiency is low in the region $>1650$~nm, however, strong PL intensity from the laser duplicate at 1704~nm allows for the visibility of the 1708~nm transition. 
Thus, Fig.~\ref{fig:TCAD_PL_fig}(C) indicates that the Ni-SiC structure can drive the \VSi into the doublet spin channel, effectively quenching the signatures of the quartet state in the affected region. 

Divacancy-related signatures were also observed in the \SI{1.3}{\micro\meter} box profile samples with $\sim$\SI{5e15}{\per\centi\meter\cubed} expected \VSi density, with PL4 being the most prominent. Interestingly, quenching of PL4 was also observed in the presence of the Ni-contact, similar to the case of \VSi, see Supplementary Results B. 

\section*{Discussion}

Opto-magnetic measurements in the form of ODMR, PL,  LE-$\mu$SR and CL spectroscopy reveal abnormal behaviour for \VSi quantum defects in 4H-SiC under Schottky contacts. By spatially mapping the \VSi luminescence signal under different contact materials, the Pd and Ni contacts are found to quench the observed signal $\sim$500~nm into the material from the contact-semiconductor interface. Schottky device simulations evidence that although the behavior under Pd metal can be explained by band bending effects, the impact of Ni on the \VSi opto-spin properties cannot, and is instead attributed to the spontaneous magnetization of Ni. 

A likely explanation for the observed magnetic contamination effects is spin channel mixing induced by stray magnetic fields from the Ni, causing the \VSi to inhabit the doublet $S=\frac{1}{2}$ instead of the commonly probed quartet channel $S=\frac{3}{2}$.  
Indeed, the 70~MHz ground state zero-field splitting of the \VSi \cite{ohshima_creation_2018, soykal_silicon_2016} corresponds to a 2.5~mT magnetic field, which means that fields exceeding this threshold have a high chance of causing strong eigenstate mixing. 
The exact mechanism and extension of this effect is left for future study. 
A possible avenue of investigation is to monitor the depth dependence of the 1708~nm \VSi doublet peak observed herein, which should behave like the inverse of the V1$^{\prime}$ signature under the Ni-contact. Further studies of contact effects on EDMR-based readout are encouraged, where an effect enhancement is expected from the close proximity of the contacts to the recombination region in which the resonance occurs.

Interestingly, the detrimental impact of Ni on \VSi\ emission shown herein likely explains the lack of visible emission enhancement under the Schottky diode studied in Ref.~\cite{Ousdal_2025}. This is because the electric field imposed by the Schottky diode at 0~V extends only $\sim300$~nm into the highly doped 4H-SiC epi-layers ($N_\mathrm{D}\sim$\SI{1e17}{\per\centi\meter\cubed}) employed. However, different biasing conditions change the electric field extension, resulting in defect emission variation as a function of applied bias, despite the emission enhancement imposed by the band bending being absent. 

The present study evidences that built-in properties, such as the spontaneous magnetization of contact materials, directly determine the performance of quantum devices and the quantum defects embedded therein. By uncovering the pronounced influence of magnetic contacts on defect spin and charge configurations, this work elevates contact engineering from a classical device consideration to a central design principle for robust quantum architectures.

\section*{Materials and Methods} 

\subsection*{Optically detected magnetic resonance measurements}
ODMR measurements were carried out at room-temperature using a custom-designed confocal fluorescence microscope setup at NASA JPL. The sample was mounted on a stripline resonator and enclosed within four layers of mu-metal shielding (Twinleaf MS-1L) to suppress external magnetic field disturbances, thereby enabling the detection of minute linewidth broadenings. Optical excitation was achieved with an 808~nm laser source (Lumics), which was coupled into an optical fiber, directed through a dichroic mirror, and focused onto the sample using a $50\times$ objective lens. The resulting photoluminescence (PL) signal was collected through the same objective and separated from the excitation path by the dichroic mirror. Residual excitation light was further attenuated using an 850~nm long-pass filter. The filtered PL signal was then routed to a fiber-coupled balanced photodetector (Thorlabs APD440A). For detection of magnetic resonance, a RF signal in the $10 - 600$ MHz range was synthesized by a Stanford Research SG382 signal generator, amplified through a Mini-Circuits ZHL-1A+ amplifier before being delivered to the stripline resonator. The RF signal was amplitude-modulated at 667~Hz, also referencing a Stanford Research SR850 lock-in amplifier, to recover the spin-dependent response fraction of the photoluminescence signal from the photodetector.

\subsection*{Low-energy muon spin rotation measurements}

Muon spin relaxation ($\mu$SR) experiments were performed with the low-energy $\mu$SR spectrometer (LE-$\mu$SR) at the $\mu$E4 beamline of the Swiss Muon Source (Paul Scherrer Institut, Switzerland) \cite{morenzoni_low-energy_2000,janka_improving_2024}. In a $\mu$SR experiment, a nearly \SI{100}{\percent} spin-polarized beam of positive muons ($\mu^+$) is implanted into the sample, where the time evolution of the muon spin polarization provides a sensitive probe of the local magnetic environment and electronic properties. In the sample, the $\mu^+$ precesses around the magnetic field (\textbf{B}) at the Larmor frequency $\omega_L = \gamma_\mu B$, with $\gamma_\mu$ being the $\mu^+$ gyromagnetic ratio. The muon decays with a mean lifetime of \SI{2.2}{\micro\second}, emitting a positron preferentially along the muon spin direction. The detection of these positrons by detectors surrounding the sample enables the reconstruction of the muon spin polarization as a function of time.
After implantation in a semiconductor, the positive muons may remain in their positive state Mu$^+$ or capture one electron to form muonium (Mu$^0$=$\mu^+ + e^{-}$), a light hydrogen‑like isotope. Details about the Mu$^0$ precession in 4H-SiC can be found in Ref.~\cite{martins_depth_2023}. A negative muonium state (Mu$^-$) may also form if a second electron capture occurs; however, this is unlikely to be detected with LE-$\mu$SR in low-doped 4H-SiC \cite{martins_depth_2023}.
Each sample, measuring $25\times 25$~\SI{}{\milli\meter\squared}, was glued onto a Ni-coated sample plate and placed on a Konti cryostat.
During LE-$\mu$SR measurements, muons are 
implanted at kilovolt energies and shallow depths (typically a few nanometers to a few hundred nanometers) to investigate local magnetic and electronic properties of materials \cite{prokscha_new_2008}. 
The LE-$\mu$SR data were recorded as a function of muon implantation energy, at a temperature of \SI{260}{\kelvin}, and in a transverse magnetic field of either \SI{0.5}{\milli\tesla} or \SI{10}{\milli\tesla}. 
The depth dependence of the muon signal was obtained by measuring as function of $\mu^+$ implantation energy between \SIrange[]{2}{22}{\kilo\electronvolt}. See Supplementary Material for simulated muon stopping profiles.

\subsection*{Cathodoluminescence measurements }

Optical characterization in the form of Cathodoluminescence (CL) was performed in a JEOL IT-300 scanning electron microscope (SEM), equipped with an integrated Delmic SPARC CL system, an Andor Shamrock SR-193i spectrograph with a 300 l/mm grating, and a charge-coupled device (CCD) Andor Newton DU940P-BU2 detector suitable for detecting luminescence in the UV/visible part of the spectrum. CL spectra were acquired at 80~K, with a probe current of $\sim1$~nA and at an accelerating voltage of 10~kV, which approximately translates to 200-700~nm penetration depth into 4H-SiC based on simulations in the software CASINO \cite{drouin_casino_2007}. Odemis software is used to control the CL mirror and the following CL data acquisition.   

\subsection*{Device simulations}
A SiC Schottky diode structure consisting of an epitaxial layer of \SI{10}{\micro\meter} was simulated with nitrogen doping of \SI{1e15}{\per\centi\meter\cubed} and substrate doping of \SI{1e19}{\per\centi\meter\cubed}. Device simulations were performed using a two-state trap model, which required the implementation of an extended Fermi-Dirac statistics for multi-state traps, considering the need to include both the ($0/-$) and ($-/2-$) charge transition levels of \VSi \cite{Hornos2011,bathen_electrical_2019}. This ensures the normalization of state probabilities, in contrast to the treatment of independent traps. 
For the device simulations, a Poisson solver has been developed, including the multi-state traps on the right hand side. The simulations were performed in equilibrium; no bias voltage was applied. With that, the Fermi level and the electron and hole quasi-Fermi energies are pinned to the resource at 0~eV. A \VSi ensemble was implemented uniformly throughout the epi-layer to a density of either \SI{5e13}{\per\centi\meter\cubed} or \SI{5e15}{\per\centi\meter\cubed}. 

\subsection*{Photoluminescence spectroscopy in the infrared}
The photoluminescence (PL) setup consists of a microscope ($10\times$ objective, Mitutoyo) coupled to an imaging spectrograph (Horriba Jobin Yvon, iHR550) equipped with a 300~grooves/mm grating. The spectrometer is coupled to an InGaAs detector array (Andor DU491A), with a spectral resolution of $\sim$0.14~nm with the 300~gr/mm grating. The sample is cooled down to cryogenic temperatures ($\sim$3.6~K) using a closed-cycle He cryostat. 
To optically excite the sample, a 852~nm continuous wave (cw) laser was used (tunable with attenuator from 0-87~mW). Using a power of 20~mW, we achieve a power density of $\sim180$~W/cm$^2$ in the beam spot. The laser excitation was directed toward the sample surface at an incident angle of 27$^\circ$ relative to the surface normal. 

\section*{Data availability}

All data needed to evaluate the conclusions in the paper are present in the paper and/or the Supplementary Materials. Source data are provided with this paper.

\section*{Acknowledgments}

Financial support was kindly provided by Akademiaavtalen between Equinor and the University of Oslo through the research project QSenS, and by the Research council of Norway through the Centre for Defects in Semiconductors for Quantum Sensing (Project No.~354831), the FRIPRO project Towards Scalable Quantum Technologies (Project No.~354419), and the Norwegian Micro- and Nano-Fabrication Facility, NorFab, Project No.~349807. The muon experiments are performed at the $\mu$E4/LEM beamline~\cite{prokscha_new_2008} of the Swiss Muon Source S$\mu$S, Paul Scherrer Institute, Villigen, Switzerland. M.M.M. and P.K.'s work is supported by the Swiss National Science Foundation under Grant No. 192218. A.G., H.K., and C.C.'s ODMR research was carried out at the Jet Propulsion Laboratory, California Institute of Technology, under contract with the National Aeronautics and Space Administration (contract 80NM0018D0004).
The work of M.E.B. was partially supported by an ETH Z{\"u}rich Postdoctoral Fellowship. 

\section*{Author contributions}
A.T.: Conceptualization, Writing—original draft, investigation, writing—review and editing, methodology, resources, data curation, validation, formal analysis, and visualization.
M.M.M.: Conceptualization, investigation, writing—review and editing, data curation, validation, formal analysis, and visualization.
A.G.: Investigation, writing—review and editing, data curation, validation, formal analysis, and visualization.
A.E.: Investigation, writing—review and editing, data curation, validation, and formal analysis.
P.K.: Methodology, writing—review and editing, and resources.
E.L.O.: Methodology, writing—review and editing, and resources.
V.B.: Methodology, writing—review and editing, and resources.
A.G.: Methodology, writing—review and editing, and resources.
T.P.: Methodology, writing—review and editing, and resources.
H.K.: Resources, writing—review and editing, and funding acquisition.
C.C.: Resources, writing—review and editing.
U.G.: Conceptualization, writing—review and editing, resources, funding acquisition, and supervision.
L.V.: Conceptualization, writing—review and editing, methodology, resources, funding acquisition, data curation, supervision, and project administration.
M.E.B.: Conceptualization, investigation, writing—review and editing, methodology, resources, funding acquisition, data curation, supervision, and project administration.

\section*{Competing interests}
The authors declare no competing interests.

\bibliographystyle{naturemag}
\bibliography{bibliography}

\newpage
\clearpage
\onecolumngrid

\makeatletter
\setcounter{secnumdepth}{3}
\makeatother
\setcounter{section}{0}
\renewcommand{\thesection}{S\arabic{section}}
\renewcommand{\thesubsection}{\Alph{subsection}}
\setcounter{figure}{0}
\renewcommand{\thefigure}{S\arabic{figure}}
\setcounter{table}{0}
\renewcommand{\thetable}{S\arabic{table}}
\setcounter{equation}{0}
\renewcommand{\theequation}{S\arabic{equation}}

\input{supplementary}

\end{document}

%% file: supplementary.tex
\begin{center}
    \large \textbf{\textit{Supplementary Material}} \\ \Large \textbf{Device contacts as spin-state selectors for silicon vacancies in 4H‑SiC}
\end{center}

\section{Supplementary Methods}
\subsection{Sample fabrication}

\subsubsection{Sample fabrication for PL and ODMR measurements} 
\begin{figure}[b]
    \centering
    \includegraphics[width=0.7\linewidth]{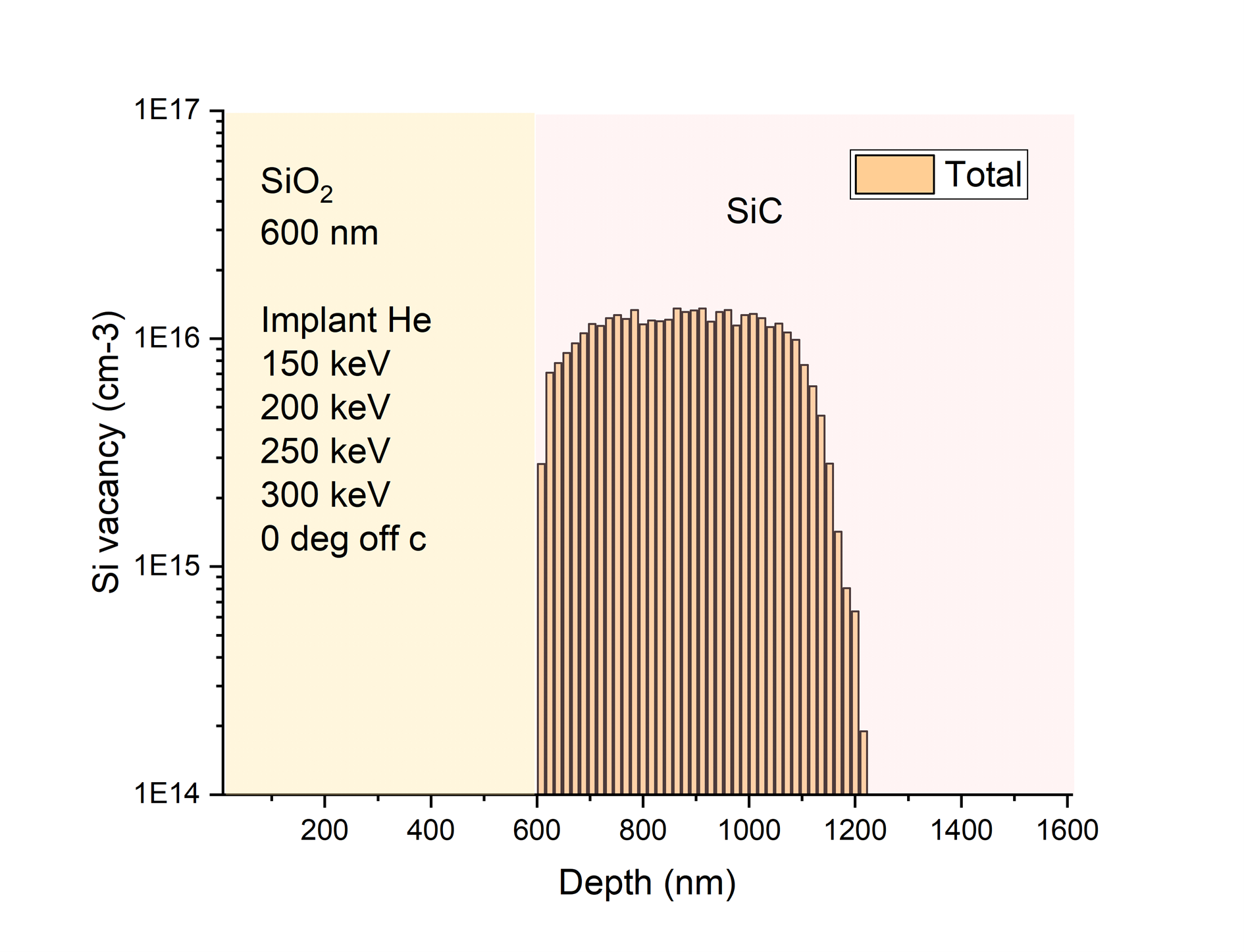}
    \caption{SRIM simulation of box profile depth of silicon vacancies predicted for different energy helium implantation to fluences of $1\times$10$^{11}$~cm$^{-3}$, yielding a predicted V$_\mathrm{Si}$ density of $\sim$\SI{1e16}{\per\centi\meter\cubed}. The yellow region is the SiO$_2$ sacrificial layer deposited by PE-CVD, that was later etched away. }
    \label{fig:odmr-samples-srim}
\end{figure}

The box profiles of Si vacancies monitored using PL and ODMR from the sample backside were fabricated using ion implantation through a sacrificial oxide capping layer. The samples under study were n-type 4H-SiC epitaxial layers having \SI{10}{\micro\meter} thickness and  $\sim2.8\times10^{15}$~\SI{}{\per\centi\meter\cubed} free carrier concentration, purchased from CREE/Wolfspeed. 600~nm of SiO$_2$ was deposited using plasma enhanced chemical vapor deposition (PE-CVD). The samples were subsequently exposed to on-axis helium implantation to four different energies to form a box profile of damage as shown in Fig.~\ref{fig:odmr-samples-srim}: 150~keV, 200~keV, 250~keV and 300~keV. The predicted depth of the box-profile is $\sim$600~nm, as simulated by simulations using the Stopping and Range of Ions in Matter (SRIM) code \cite{ziegler_srim_2010}. SRIM accurately predicts the formation depth of $\text{V}_\text{Si}$ in SiC~\cite{krausThreeDimensionalProtonBeam2017}. Two different sample sets were formed using different fluences of the helium beam, \SI{1e9}{\per\centi\meter\squared} and \SI{1e11}{\per\centi\meter\squared}. Thereafter, the oxide layer was etched away in HF, and the samples were exposed to a \SI{300}{\celsius} annealing step for 30~min in N$_2$ flow to alleviate implantation damage. This yielded predicted V$_\mathrm{Si}$ densities of \SI{1e14}{\per\centi\meter\cubed} and \SI{1e16}{\per\centi\meter\cubed}, respectively, when assuming $3 \ \%$ of  V$_\mathrm{Si}$ survive dynamic annealing during implantation and the \SI{300}{\celsius} post-implantation heat treatment \cite{bathen_electrical_2019}. Finally, the samples were diced, and two sets of Schottky diodes were the result: one with a thin layer (20~nm) of metal, and one with a thick layer (100~nm) of metal. The metals employed were Al, Ti, Ni and Pd.

\subsubsection{Sample fabrication for LE-$\mu$SR measurements} 
We employed n-type 4H-SiC (0001) samples with epi-layers of \SI{10}{\micro\meter} thickness and  $\sim2.8\times10^{15}$~\SI{}{\per\centi\meter\cubed} free carrier concentration, purchased from CREE/Wolfspeed and diced into 2.5$\times$2.5 \SI{}{\centi\meter\squared} pieces. Si vacancies were formed by proton irradiation performed at the University of Oslo with 1.8~MeV energy and $1\times10^{14}$~\SI{}{\per\centi\meter\squared} fluence, further followed by post-irradiation annealing at \SI{300}{\celsius} for 30~min in N$_2$ flow. This is done to minimize the immediate implantation damage such as interstitial defects. After the defect creation, different metals of 20~nm thickness were deposited on the entire sample surface using electron beam evaporation. Three different metals were employed as Schottky contacts on the 4H-SiC epi-layers, Ni, Pd, and Ti, as 20~nm Al layers have previously been shown to have no impact on the muon signal in 4H-SiC \cite{Kumar_2023}. 

\subsubsection{Sample fabrication for CL measurements}\label{sepMethod:CL}
\begin{figure}[h]
    \centering
    \includegraphics[width=\linewidth]{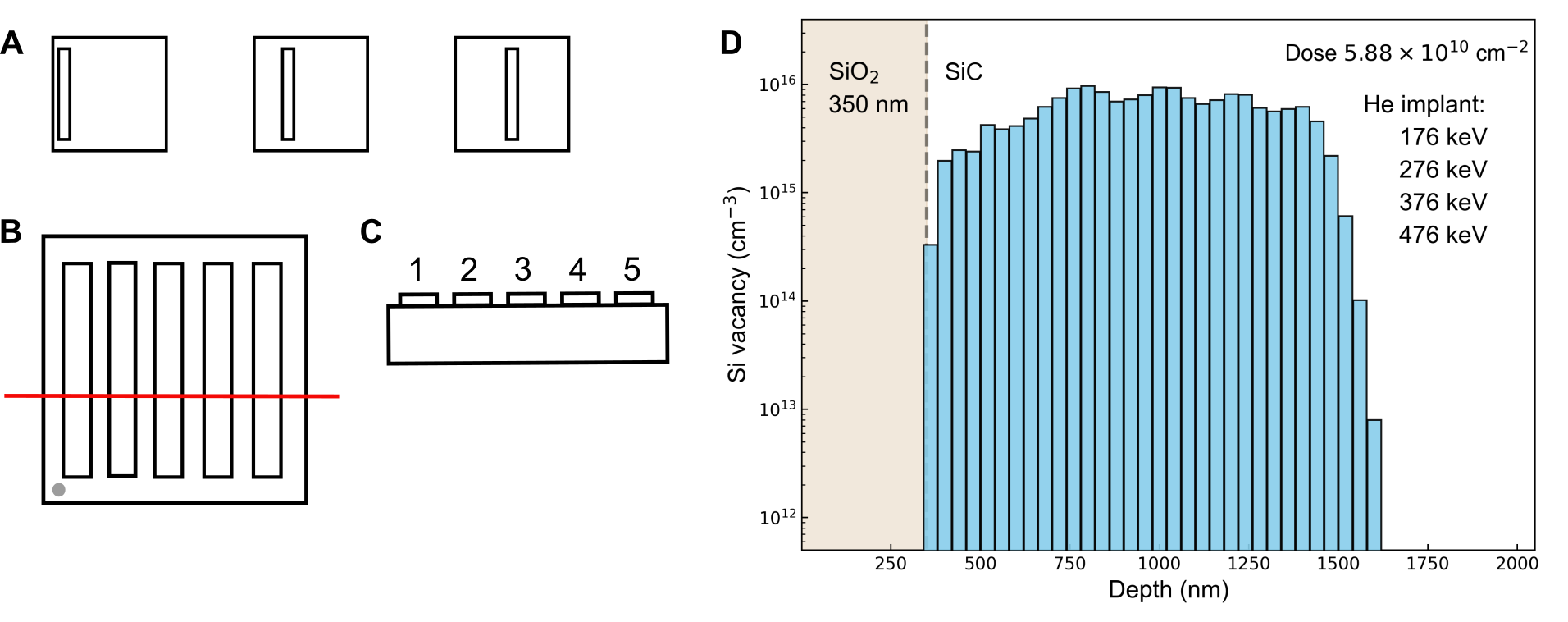}
    \caption{(A) Masks for metal deposition, specifically designed for making contacts in stripes of the individual five materials, making up the final structure schematically drawn in (B). The red line visualizes the crossectional cut, viewed from the side in (C). Numbered from 1-5, from left to right are the materials Au, Ti, Ni, Al, and Pd, respectively. (D) SRIM simulation of box profile depth of silicon vacancies for \SIrange{176}{476}{\kilo\electronvolt} He implantation to a fluence of 5.62$\times$10$^{10}$~cm$^{-2}$. The beige region is the PVD-deposited SiO$_2$ sacrificial layer, that was later etched away. }
    \label{fig:masks_SRImbox}
\end{figure}

An n-type 4H-SiC wafer holding a \SI{10}{\micro\meter} thick epitaxial layer with free carrier concentration of 1$\times$10$^{15}$~cm$^{-3}$ was cut into $7\times7$ mm$^2$ squares and implanted with He$^+$ by ion implantation. The He ions were implanted at fluences of 5.62$\times$10$^{10}$~cm$^{-2}$ (samples A) or 5.62$\times$10$^{11}$ cm$^{-2}$ (samples B). Energies of 176~keV, 276~keV, 376~keV, and 476~keV were employed to obtain a box profile with defect concentrations of 5$\times$10$^{15}$ and 5$\times$10$^{16}$~cm$^{-3}$, respectively (see Fig.~\ref{fig:masks_SRImbox}(D)). A 350~nm thick layer of SiO$_2$ was deposited onto the surface of SiC as a sacrificial layer prior to ion implantation, visible as the beige square in Fig.~\ref{fig:masks_SRImbox}(D). This deposition was performed with a Tornado E-beam PVD system. 
The thickness of the SiO$_2$ layer was confirmed by ellipsometer and profilometer measurements. The SiO$_2$-layer was later etched away using HF for 3~min. The samples employed for CL measurements were not annealed after the implantation. 

To form distinct regions where the V$_\mathrm{Si}$ defects were exposed to the impact of different metals, stripes of different metals were deposited using a Tornado E-beam PVD, with masks of Si cut with an ELAS laser cutter, see Fig.~\ref{fig:masks_SRImbox}(A). The samples were designed to have five different parallel metal-stripes of thickness 100~nm, thus only a single stripe of metal was deposited on each sample at a time, while the rest of the sample was covered by the mask. 
The metals used were Au, Ti, Ni, Al, and Pd, numbered 1-5 respectively in Fig.~\ref{fig:masks_SRImbox}(C). Of these, Au, Ni, Ti, and Al are commonly used as contact materials on SiC, however, Al is more common to use on p-type SiC. Pd is included because it is paramagnetic, and hence may exert an influence on the electro-magnetic environment experienced by V$_\mathrm{Si}$ in 4H-SiC. All metals employed herein are expected to result in a Schottky-type contact on the n-type SiC because of their high work functions (see Table~\ref{tab:electricalStats}). The Au-contact exhibited bad adhesion to SiC and thus no distinct defect signatures were observed in the CL measurements from beneath the Au contact. Therefore, Au-related measurements will not be discussed further herein. 

Prior to CL characterization, cross-sections of the samples were fabricated with an ELAS laser cutter. The cross-sectional cut is shown in Fig.~\ref{fig:masks_SRImbox}(B), and the schematic cross-sectional profile is shown in Fig.~\ref{fig:masks_SRImbox}(C). The materials Au, Ti, Ni, Al, and Pd are numbered 1-5 in the figure.

\subsection{LE-$\mu$SR data analysis}

\begin{figure}[h!]
    \centering
    \includegraphics[width=0.45\linewidth]{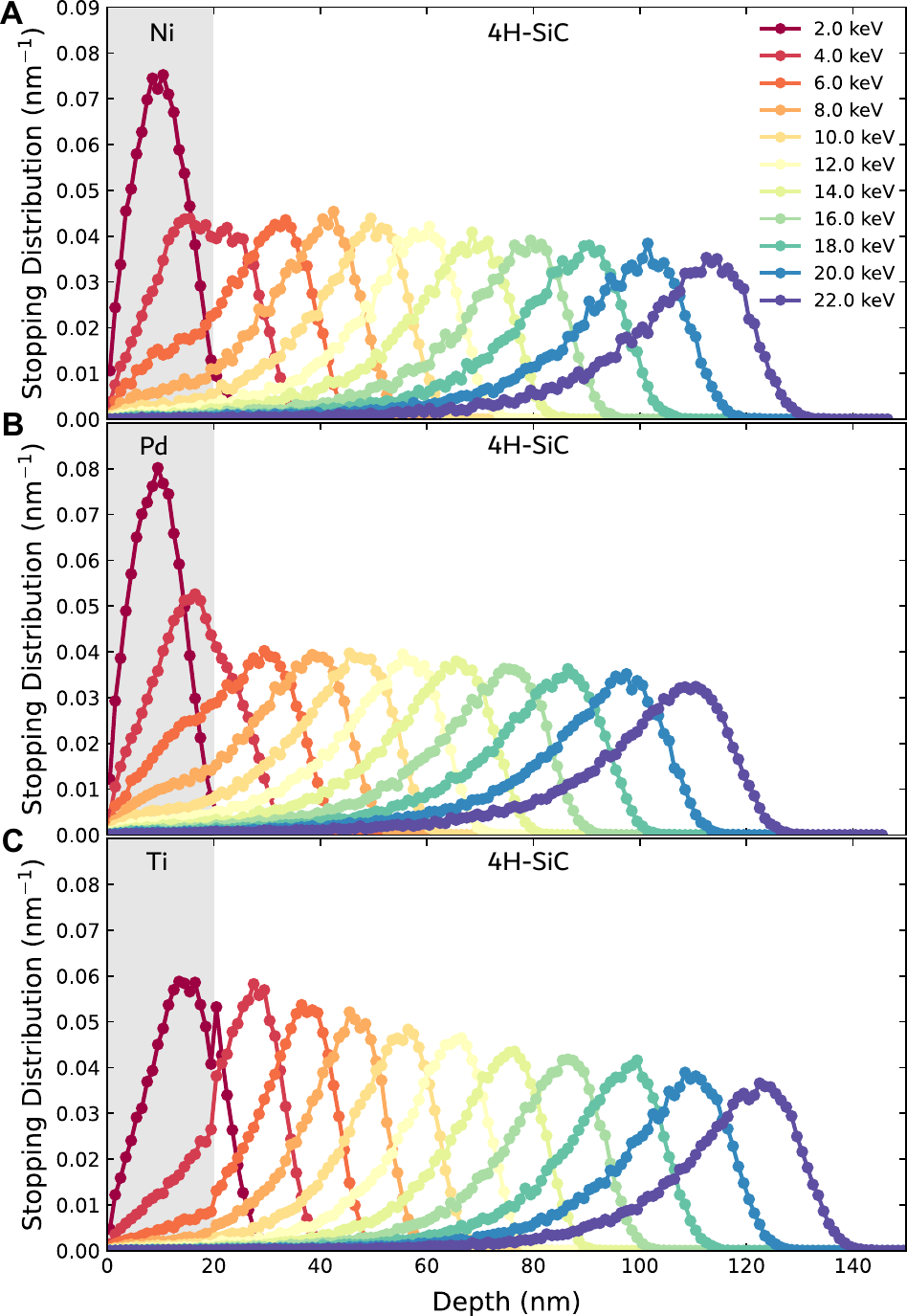}
    \caption{Stopping distribution of the muon beam implanted with energies between \SIlist{2;22}{\kilo\electronvolt}. The profiles are obtained with Monte Carlo simulation TRIM.SP \cite{Morenzoni_2002,Eckstein_1991}.}
    \label{fig:uSR-stopping}
\end{figure}

This section addresses the analysis of the LE-$\mu$SR data presented in Fig.~2 of the main text, and the simulations of the stopping distribution of the muons, required to extract depth-dependent information.
Fig.~\ref{fig:uSR-stopping} shows the predicted stopping profiles of different energy muons throughout the metal/SiC stack for Ni (Fig.~\ref{fig:uSR-stopping}A), Pd (Fig.~\ref{fig:uSR-stopping}B) and Ti (Fig.~\ref{fig:uSR-stopping}C) as contact metal simulated with TRIM.SP \cite{Morenzoni_2002,Eckstein_1991}. Note that the Al/SiC sample was not measured with LE-$\mu$SR due to the limited allocated beamtime. However, previous measurements on Al/SiO$_2$/SiC structures reveal no impact of the metal on the $\mu$SR signal, when compared to a sample without Al \cite{Kumar_2023}.

All the LE-$\mu$SR data were analyzed using musrfit \cite{suter2012musrfit}. 
In metallic layers, implanted muons remain predominantly in the diamagnetic $\mu^+$ state and give rise to a single precession signal. In contrast, in SiC the implanted muons can capture an electron to form neutral muonium (Mu$^0$), resulting in the coexistence of diamagnetic and paramagnetic states \cite{martins2026muoniumdynamicsprobedepthresolved}. Consequently, two distinct components are typically observed in the LE-$\mu$SR spectra. The time spectra were therefore fitted using a model consisting of diamagnetic and paramagnetic contributions whenever muonium formation was observed.
The main component can be described by a Lorentzian-damped cosine function of the form
\begin{equation}
       A_\text{Dia}(t) = A_\text{Dia} \cdot e^{(-\lambda_\text{D} \cdot t)} \cdot \cos(2\pi\nu_D t ), 
     \label{eq:AsymDia}
\end{equation}

where $A_{\rm Dia}$ is the asymmetry amplitude of the diamagnetic component (due to formation of a charged state Mu$^+$ or Mu$^-$), $\lambda_{\rm D}$ is the corresponding depolarization rate, and $\nu_{\rm D}$ is the muon spin precession frequency. The phase $\phi_\textrm{D}$ of the diamagnetic signal is usually the angle offset of the decay positron detector in relation to the muon spin polarization at $t=0$.
To account for the Mu$^0$ formation in the 4H-SiC epi-layer observed at \SI{0.5}{\milli\tesla}, an additional contribution comprising two components was included,
\begin{equation}
       A_\text{Mu}(t) =  A_\text{Mu} \cdot e^{(-\lambda_\text{Mu}\cdot t)}\cdot [\cos(2\pi\nu_1 t + \phi_\text{Mu}) + \cos(2\pi\nu_2 t + \phi_\text{Mu})],
     \label{eq:AsymMu}
\end{equation}
where $A_{\rm Mu}$ is the asymmetry amplitude associated with the paramagnetic fraction, accounting for both frequency lines. $\lambda_\text{Mu}$ is the exponential depolarization rate of Mu$^0$, and  $\nu_{1/2}$ are the precession frequencies of Mu$^0$ at low magnetic field.
 
The data collected at \SI{10}{\milli\tesla} was fitted using only the first component of Eq.~\ref{eq:AsymDia}, since only the diamagnetic contribution is observable in the LE-$\mu$SR spectrometer at this field.

When Mu$^0$ was observed in 4H-SiC at \SI{0.5}{\milli\tesla}, the time spectra was fitted as:
\begin{equation}
       A(t) = A_\text{Dia} (t) + A_\text{Mu} (t).
     \label{eq:Asy_t}
\end{equation}
The asymmetry parameter  $A_\text{D}$ was converted into the diamagnetic fraction F$_\text{D}$, by normalizing to the maximum asymmetry measurable in silver (Ag), where the muon spin depolarization is very weak. The fraction is calculated as F$_\text{D}=A_\text{D}/A_\text{Ag}$.
More details about fitting the LE-$\mu$SR recorded in 4H-SiC can be found in Refs.~\cite{martins_depth_2023,Kumar_2023}. 

\subsection{CL data processing}
\begin{figure}[h]
    \centering
    \includegraphics[width=0.9\linewidth]{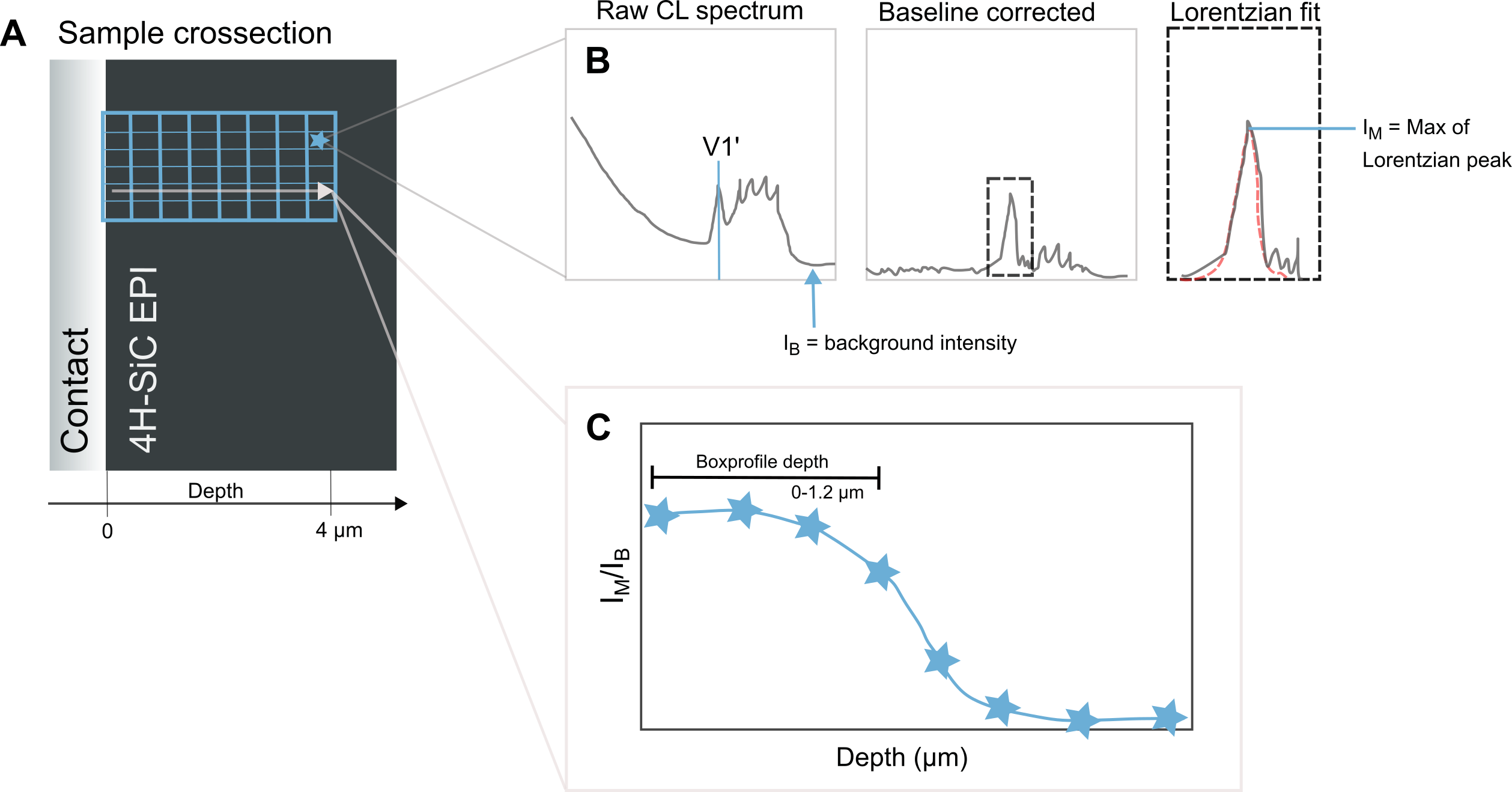}
    \caption{Overview schematic of CL experiment. (A) Sample cross-section with 2D measurement grid where each pixel (single pixel marked with blue star) contains a luminescence spectrum as seen in (B). (B) Data processing procedure of the luminescence spectrum. The V1$^{\prime}$ peak is highlighted with blue line. The spectrum is further baseline corrected and V1$^{\prime}$ is fit with a Lorentzian function to extract the maximal intensity I$_M$. (C) Depth profile of V1$^{\prime}$ luminescence, normalised as I$_M$/I$_B$ where I$_B$ is the background intensity. Each blue star corresponds to the pixels along the grey arrow in (A) highlighting the row from where the data originates. }
    \label{fig:dataProcessingCL}
\end{figure}

CL data is initially examined with Odemis Viewer (the official Odemis software for visualizing data), before further post-processing was performed with Python. The output data from CL consists of an overview SEM image of the area inspected, and a 2D grid of pixels where each pixel contains a luminescence spectrum of intensity as a function of wavelength, see Fig.~\ref{fig:dataProcessingCL}(A). For every geographical spot chosen for studying, we obtain two luminescence spectra of central wavelength 600~nm and 860~nm, respectively. 

The 2D grid can be tailored in size for specific needs. In this experiment, we chose a rectangle of height 6 rows, and width 37 columns, since we want to know how the intensity of defect luminescence varies as a function of depth from the surface of the sample. For more accurate statistics, we average the intensities over all rows. We are interested in the first few micrometers under the sample surface, where we expect our box-profile, seen in Fig.~\ref{fig:masks_SRImbox}(D), to be located. 
Thus, the data is obtained from an area reaching from the outside the surface (the one with contacts deposited) to a few micrometers ($\sim$4-\SI{5}{\micro\meter}) beneath the surface. A pixel-correction is necessary for all data to have the same 0-point for where the sample surface starts. This pixel-correction is done manually by observation of the SEM-image contrast between surface and bulk sample in combination with observations of where the intensity-signal of the tail around 650~nm (in the 860~nm spectra) starts. 

The CL data processing procedure is visualized in Fig.~\ref{fig:dataProcessingCL}(B). 
The goal of the processing is to map the height of V1$^{\prime}$ (ZPL of V$_\mathrm{Si}$ at hexagonal lattice site) and plot it as a function of depth into the sample, as seen in Fig.~\ref{fig:dataProcessingCL}(C). 
First, the Python-script (that can be shared upon request) converts the raw CL data from hdf5-format to csv files for each row. Using the converted csv-files, the script iterates through each of the pixels (containing a luminescence spectrum) for each row, removes extreme outliers like cosmic rays, fits the spectrum, and finally removes the baseline of the data. Further on, a Lorentzian function is fit to the V1$^{\prime}$ ZPL, and the maximum of the Lorentzian peak is extracted. This maximal intensity is normalised to the background of the data to adjust for measurement-specific variations like focus, alignment, and internal material differences.  

\subsection{Device simulation}
For the solution of the Poisson equation, including multi-state traps, a corresponding solver has been developed and used for the calculation of the occupation of the charge states of the Si vacancy in 4H-SiC (ground state: 0, singly negative: -, doubly negative: 2-). The right hand side of the Poisson equation contains the charge density from electrons, holes, doping, and singly and doubly charged defect states. The sum of the occupation probabilities of all states has to be equal to one, which connects the state occupation probabilities via a Grand Partition Function. These state occupation probabilities go into the right hand side of the Poisson equation as a weight function of the defect concentrations.

\newpage
\section{Supplementary Results} 
\begin{figure}[h]
    \centering
    \includegraphics[width=0.8\linewidth]{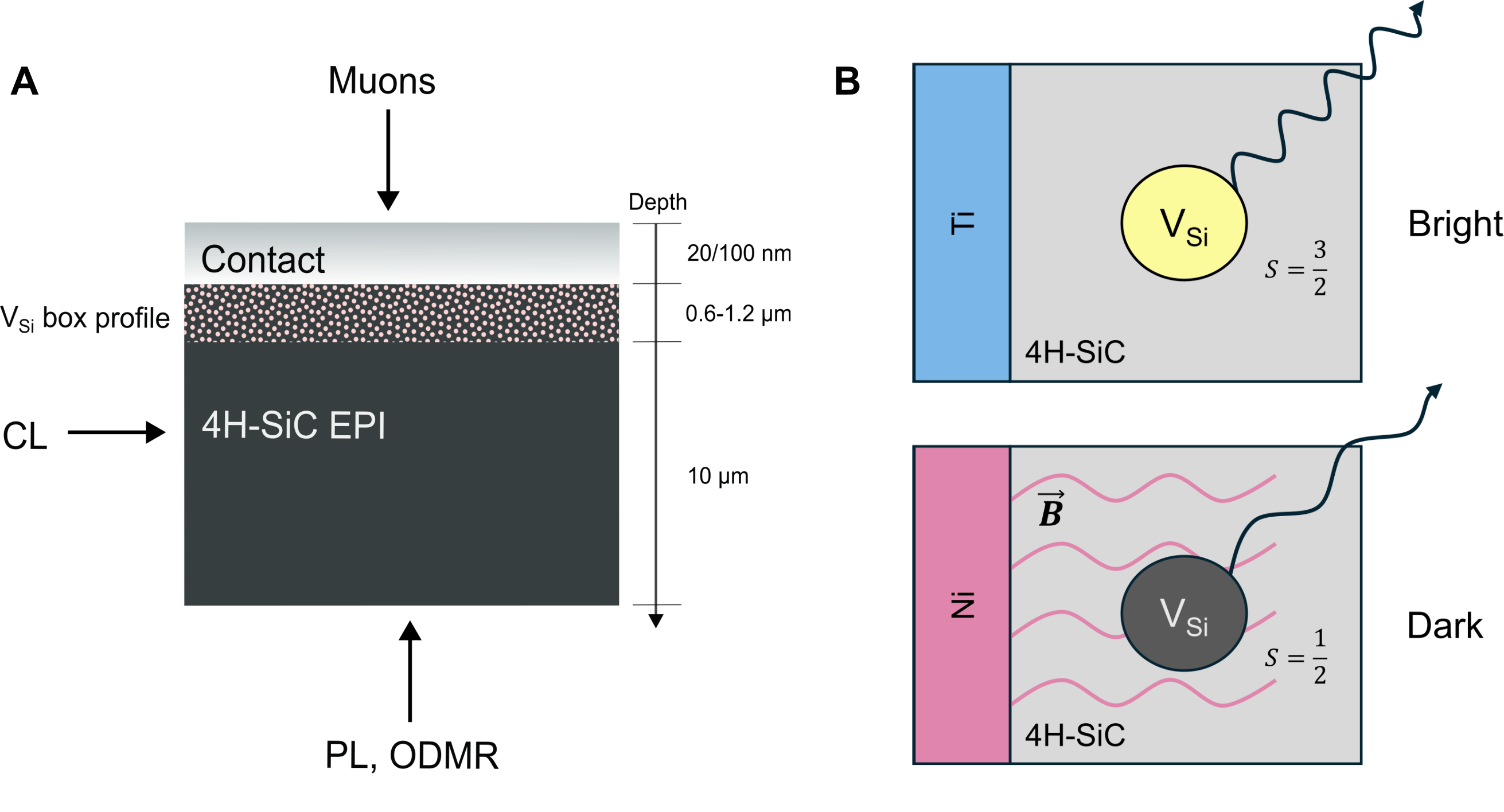}
    \caption{(A) Sample geometry with near-constant V$\mathrm{_{Si}}$ concentration in the upper 4H-SiC epi-layer and directions of measurement collection for different techniques. (B) Simplified schematic of the defect emission from under different contacts.  }
    \label{fig:schematic}
\end{figure}

Fig.~\ref{fig:schematic}(A) shows the measurement geometry for the different experimental methods in relation to the samples employed, where LE-$\mu$SR is performed through the metal contact, CL measurements are conducted in cross-section, and PL/ODMR signatures are collected from the sample backside. Note that different defect densities and distributions are required for each technique.
A simplified schematic of the mechanisms governing the emission of the defects is shown in Fig.~\ref{fig:schematic}(B).

\subsection{Electrical measurements of contacts on CL-samples}
\begin{figure}[t]
     \centering
     \includegraphics[width=0.9\textwidth]{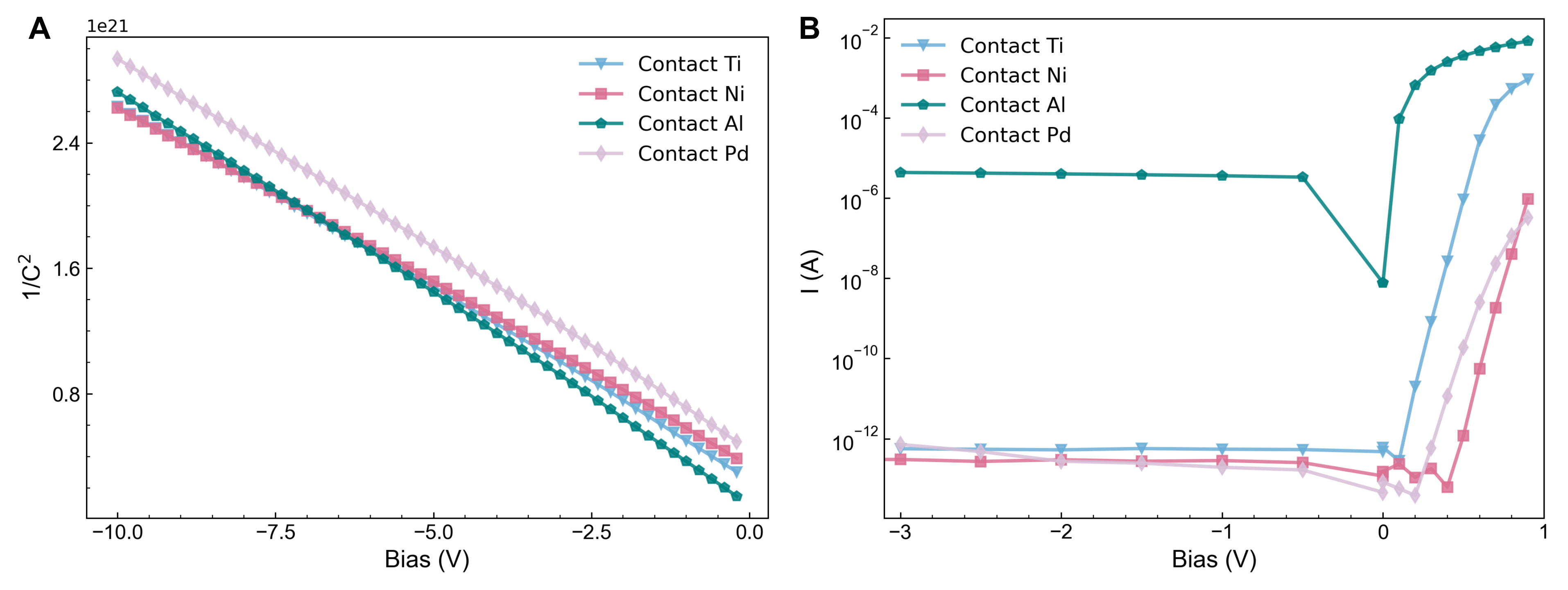}
     \caption{(A) CV and (B) IV curves of SiC Schottky diodes with contacts of different metals: Ti (blue), Ni (pink), Al (green), and Pd (light purple).}
     \label{fig:AllCVIV}
\end{figure}

Capacitance-Voltage (CV) and Current Voltage (IV) measurements were performed on similar 4H-SiC samples to those used in the CL experiment to verify the diode properties of the contacts. 100~nm thick circular contacts of diameter 1~mm were deposited, on unimplanted samples from the same SiC wafer as in the CL-experiment, for each metal in parallel with the previous metal stripe depositions performed in the Tornado e-beam PVD. The electrical measurements resulted in the (A) CV and (B) IV curves shown in Fig.~\ref{fig:AllCVIV}. All contacts exhibit rectifying behavior when exposed to applied voltage, hence they are all characterized as Schottky contacts. The net doping concentrations, built-in voltages, and barrier heights $\Phi_B$ are calculated from a linear fit of 1/$C^2$ as a function of applied bias, see Table~\ref{tab:electricalStats}. The barrier height values are comparable with previous findings \cite{itoh_high_1995, harrell_aluminum_2002, zekentes_schottky_2018}, and the measured doping concentrations are similar for all contacts.

The linear regions of the IV-curves for each contact metal were fitted with a linear regression model to find the ideality factor as the slope of linear region. The calculated ideality factors of all contacts are presented in Table \ref{tab:electricalStats}. The experimental data from the Al-contact did not exhibit a linear region as the other contacts did, and was therefore not possible to fit. 

\begin{table}[h]
\centering
\caption{Electrical measurement results for contacts Ti, Ni, Al, and Pd. Ideality factors are extracted from IV measurements, while net doping concentrations, built-in voltage V$_{bi}$, and barrier height $\Phi_B$ are determined from CV measurements. }\label{tab:electricalStats}
{\setlength\tabcolsep{8pt}
\begin{tabular}{ccccc}\toprule
 Material & Ideality factor & Net doping conc. (\SI{}{\centi\meter\cubed})& V$_{bi}$ (eV) & $\Phi_B$ (eV)\\
\midrule 
 Ti & 1.05 &$1.0\times10^{15}$& 1.20 & 1.45\\
 Ni & 1.15 &$1.0\times10^{15}$&1.63 & 1.88 \\
Al & - &$9.0\times10^{14}$& 0.47 & 0.72 \\
 Pd & 1.38 &$9.6\times10^{14}$& 1.95 & 2.20 \\ 
 \botrule
\end{tabular}}

\end{table}


\subsection{ODMR on additional samples}
\begin{figure}[h]
    \centering
    \includegraphics[width=0.9\linewidth]{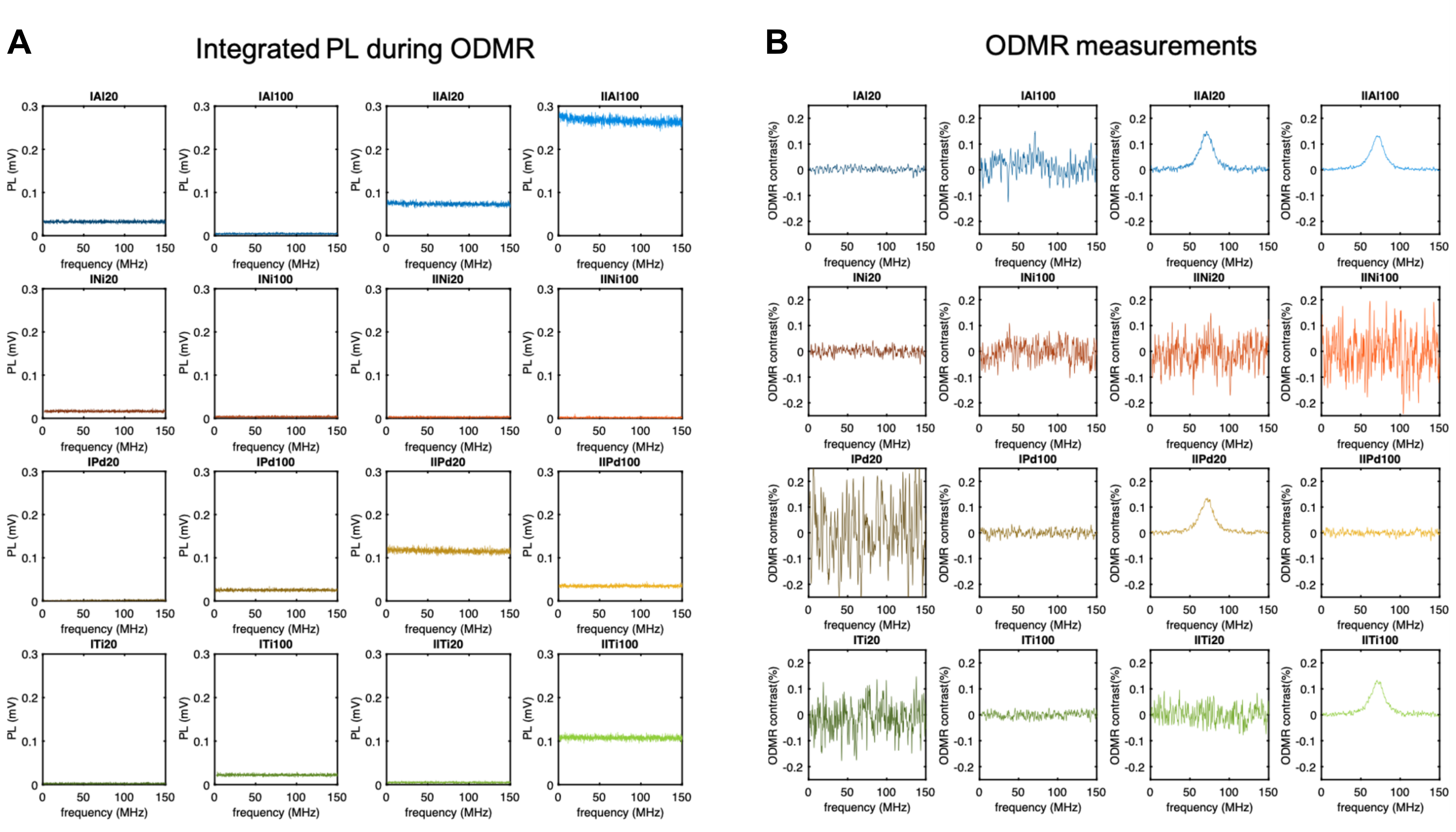}
    \caption{(A) Integrated PL during ODMR and (B) ODMR signatures for samples with V$_\mathrm{Si}$ concentrations of 10$^{14}$~\SI{}{\per\centi\meter\cubed} and 10$^{16}$~\SI{}{\per\centi\meter\cubed}, marked by I and II, respectively, and having contact thicknesses of 20 and 100~nm, marked by 20 and 100, respectively.  }
    \label{fig:PL_ODMR_sup}
\end{figure}

ODMR was performed on samples containing different V$_\mathrm{Si}$ concentrations (I: 10$^{14}$~\SI{}{\per\centi\meter\cubed} and II: 10$^{16}$~\SI{}{\per\centi\meter\cubed}), different contact thicknesses (20 and 100~nm), and different Schottky contact metals (Al, Ni, Pd and Ti). Figure~\ref{fig:PL_ODMR_sup} shows (A) the integrated PL during ODMR measurements, and (B) the ODMR signal centered on the ZFS of the $\mathrm{V_{Si}}^-$. Interestingly, we find variations depending on both the defect density (most prominent signal for higher concentration samples, labelled II), contact thickness (thicker contact could yield more reflection given the backside-collection measurement geometry), and contact metal type (assigned to a combination of band bending and metallic contamination). Indeed, $\mathrm{V_{Si}}$-related signals are hardly visible or completely gone for both PL and ODMR in the Ni-samples. 

\subsection{ODMR linewidth}
\begin{figure}[h]
    \centering
    \includegraphics[width=0.75\linewidth]{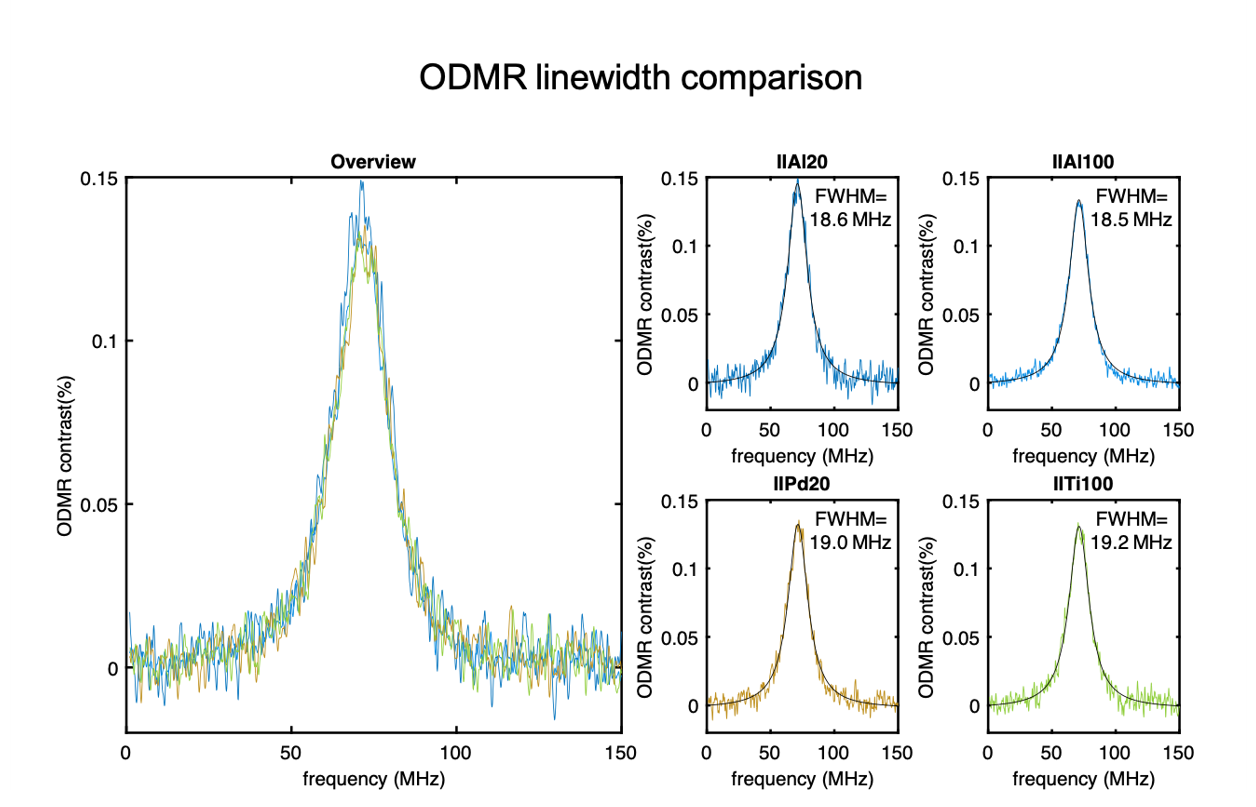}
    \caption{ODMR linewidth from selected contacts. }
    \label{fig:ODMRlinewidth}
\end{figure}

Fig.~\ref{fig:ODMRlinewidth} shows the $\mathrm{V_{Si}}$ ODMR signal for the high defect-concentration Al (20 and 100~nm thickness), Pd and Ti samples. Ni-samples displayed no visible $\mathrm{V_{Si}}$ ODMR signal. The Al-samples both have similar full width at half maximum (FWHM) linewidths of 18.6~MHz (20 nm) and 18.5~MHz (100 nm). However, the other metals, Pd and Ti, display slightly broadened linewidths of 19.0~MHz (20 nm thick contact) and 19.2~MHz (100 nm thick contact), respectively. This evidences that Al is likely the most beneficial contact metal from an ODMR perspective, however, with less beneficial Schottky contact properties. 

\newpage
\subsection{Depth Evolution of Muonium Formation}
\begin{figure}[h!]
    \centering
    \includegraphics[width=0.90\linewidth]{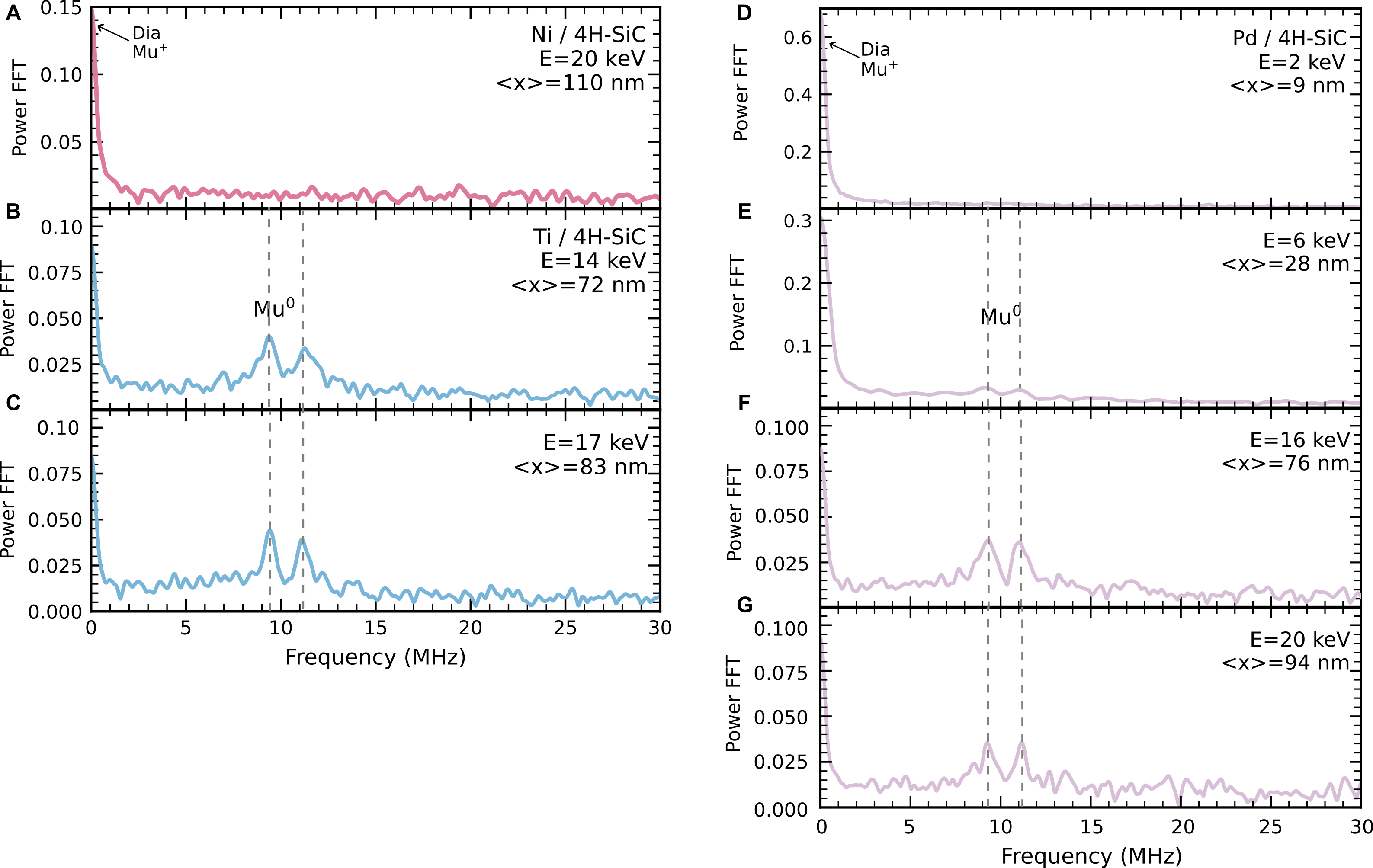}
    \caption{Power Fourier transforms of the time spectra measured with LE-$\mu$SR. The Fourier spectra are shown for Ni/4H-SiC (A), Ti/4H-SiC (B-C) and Pd/4H-SiC (D-G) samples at different implantation energies, corresponding to the indicated mean muon stopping depths $\langle x\rangle$. The dominant low-frequency line originates from the diamagnetic component, while the pair of peaks at $\sim9.8$~MHz and $\sim11$~MHz, highlighted by the dashed lines, is assigned to neutral muonium Mu$^0$. In the Ni/4H-SiC sample, the Mu$^0$ signal is observed in the SiC layers of the Ti- and Pd-samples. In contrast, the Mu$^0$ signal is not observable in the Ni/4H-SiC sample, likely due to enhanced depolarization induced by stray magnetic fields from the Ni layer.}
    \label{fig:uSR-MuLines}
\end{figure}

The Fourier transform of the time spectra recorded at \SI{0.5}{\milli\tesla} are shown in Fig.~\ref{fig:uSR-MuLines}. 
Although the measurement energy range was similar for all samples, the mean probing depth differs due to the difference in the density of the metallic layer.

As can be seen in Fig.~\ref{fig:uSR-MuLines}, no Mu$^0$ contribution is observed in the SiC layer of the sample with Ni, even at the maximum probing depth. This contrasts with the Ti and Pd samples where Mu$^0$ is undisturbed in the SiC layer. Particularly, one can see that below the Pd/SiC interface (Fig.~\ref{fig:uSR-MuLines}~(E)), the Mu$^0$ state starts to appear, revealing that Pd has no significant impact on the magnetic environment of the epi-layer. The amplitude of the Mu$^0$ signal is reduced compared to that observed at larger implantation depths (Fig.~\ref{fig:uSR-MuLines}(F-G)). This reduction is attributed to the non-negligible fraction of muons stopping in the Pd overlayer at an implantation energy of \SI{6}{\kilo\electronvolt}, thereby decreasing the number of muons reaching the SiC epilayer and contributing to the Mu$^0$ signal.

\newpage
\subsection{Device Simulation}
\begin{figure}[h]
    \centering
    \includegraphics[width=0.85\linewidth]{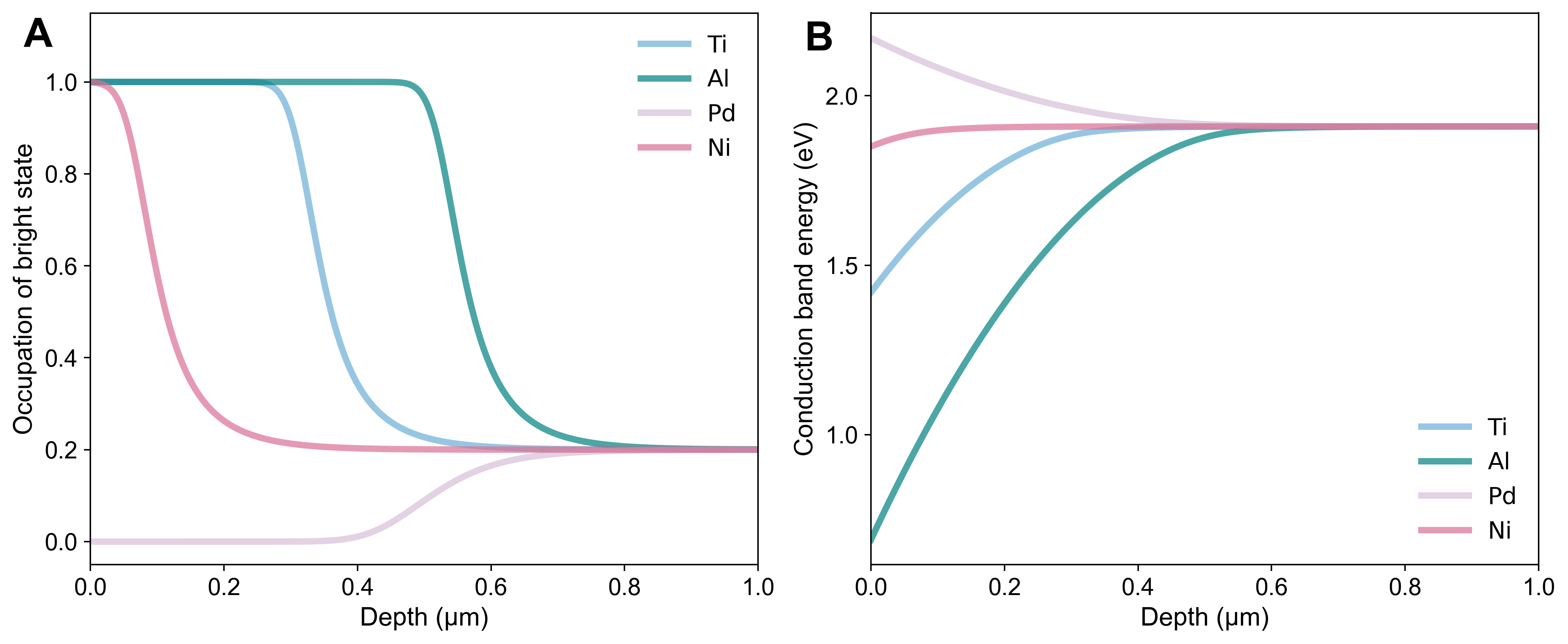}
    \caption{(A) Device simulation of the singly negative charge state occupation (energy state at 1.9~eV below the conduction band edge) of the $\mathrm{V_{Si}}$ for the contact materials Ti, Al, Pd, and Ni. The simulation was performed using a multi-state acceptor trap with energies of 1.9~eV (-) and 0.6~eV (2-) from the conduction band edge, a defect concentration of \SI{5e15}{\per\centi\meter\cubed}, at a temperature of 80~K. (B) Corresponding simulated conduction band energy as a function of depth into the material for the different contact metals.}
    \label{fig:tcad5e15}
\end{figure}

Fig.~\ref{fig:tcad5e15} shows device simulations for the high defect concentration case. Here, 
Fig.~\ref{fig:tcad5e15}(A) singly negative charge state occupation (energy state at 1.9~eV below the conduction band edge) of the $\mathrm{V_{Si}}$ for the contact materials Ti, Al, Pd, and Ni and using a defect concentration of \SI{5e15}{\per\centi\meter\cubed}. Figure~\ref{fig:tcad5e15}(B) shows the corresponding conduction band energies. Due to the high concentration of charged defects, the occupation of the energy states determines the conduction band bending. The simulations were performed with a temperature of 80~K.

\newpage
\subsection{Divacancy PL emission quenching}

\begin{figure*}[h]
    \centering
    \includegraphics[width=0.9\linewidth]{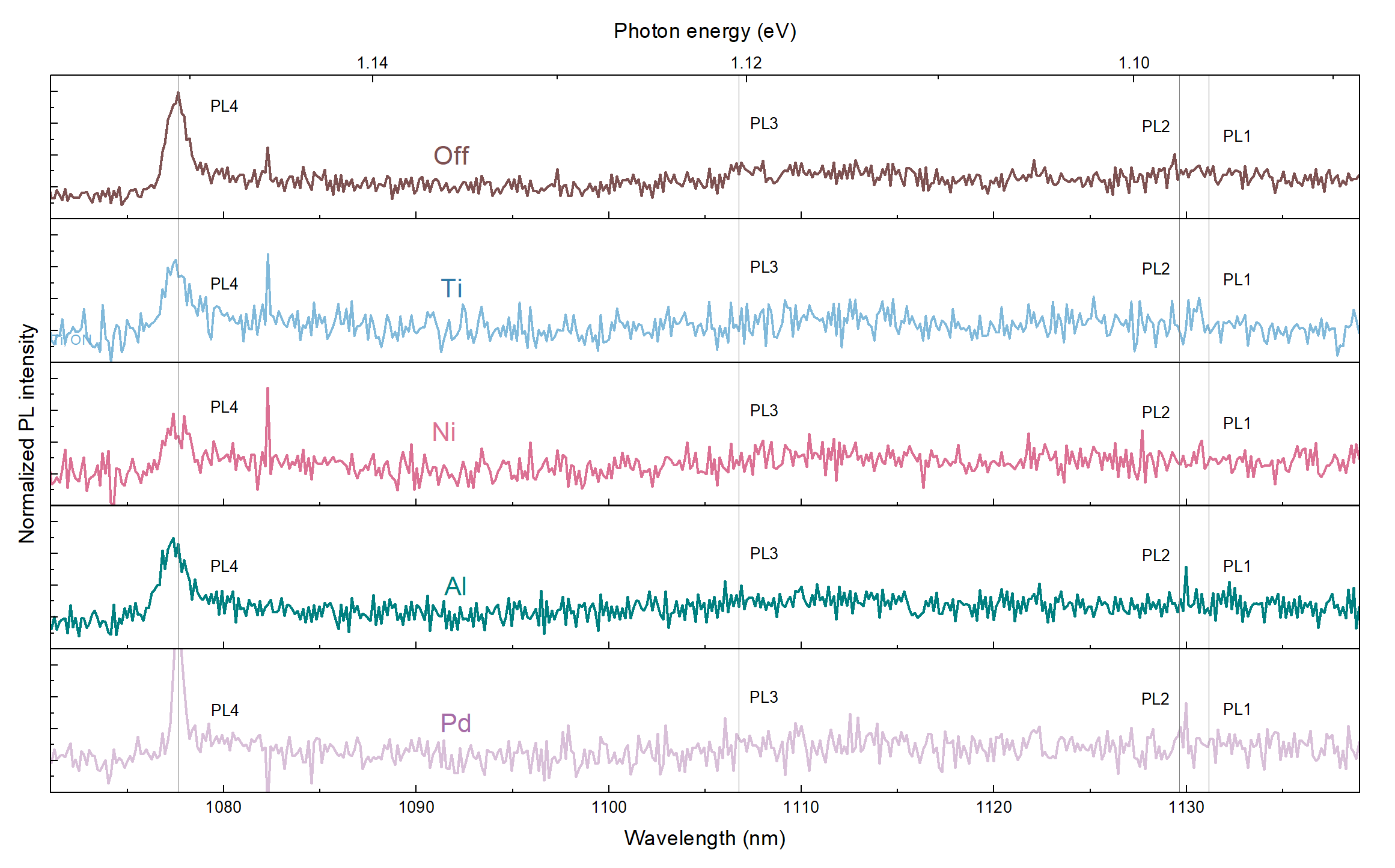}
    \caption{PL spectra of divacancy signals under all contacts. The PL intensity is normalized, to highlight the quenching of PL4 under the Ni contact. Measurement temperature is 3.6~K and excitation source 400~nm is used. }
    \label{fig:PL_PL4}
\end{figure*}

During PL-measurements, emission from the neutral divacancy (V$_\mathrm{Si}$V$_\mathrm{C}^0$) was visible under all contacts. Interestingly, the signal from the most prominent divacancy feature (PL4) is seemingly weaker under the Ni-contact than the other contacts, see Fig.~\ref{fig:PL_PL4}. This result indicates that the quenching effect from the Ni-contact extends beyond the test case of the V$_\mathrm{Si}$ alone, thus having relevance for the spin environment of quantum defects in general. Further work is needed to determine the extension of this effect including ODMR measurements for the divavancy. 

